\documentclass[12pt, preprint]{aastex} 
\usepackage{graphicx}
\usepackage{epstopdf}

\shorttitle{Spitzer power-law galaxies in EGS}
\shortauthors{Park et al.}

\newcommand{\0}{\phantom{0}}

\begin{document}

\title{AEGIS: A Multi-wavelength Study of {\it Spitzer\/} Power-law Galaxies}
\author{S.~Q.~Park\altaffilmark{1},
        P.~Barmby\altaffilmark{2,1}, 
	S.~P.~Willner\altaffilmark{1},
	M.~L.~N.~Ashby\altaffilmark{1},
        G.~G.~Fazio\altaffilmark{1},
	A.~Georgakakis\altaffilmark{3},
	R.~J.~Ivison\altaffilmark{4,5},
	N.~P.~Konidaris\altaffilmark{6},
	S.~Miyazaki\altaffilmark{7}
        K.~Nandra\altaffilmark{8},
	D.~J.~Rosario\altaffilmark{6}
        }

\altaffiltext{1}{Harvard--Smithsonian Center for Astrophysics, 60
 Garden Street, 
 Cambridge, MA 02138, USA; spark@cfa.harvard.edu}
\altaffiltext{2}{Department of Physics \& Astronomy, University of
 Western Ontario, London, ON N6A 3K7, Canada}
\altaffiltext{3}{National Observatory of Athens, I. Metaxa \&
 V. Paulou, Athens 15236, Greece} 
\altaffiltext{4}{UK Astronomy Technology Centre, Science and Technology Facilities
Council, Royal Observatory, Blackford Hill, Edinburgh EH9 3HJ, UK}
\altaffiltext{5}{Institute for Astronomy, University of Edinburgh, Blackford Hill,
Edinburgh EH9 3HJ, UK}
\altaffiltext{6}{UCO/Lick Observatory, University of California Santa
 Cruz, 1156 High Street, Santa Cruz, CA 95064}
\altaffiltext{7}{National Astronomical Observatory of Japan, Osawa
 Mitaka Tokyo 181-8588, Japan}
\altaffiltext{8}{Astrophysics Group, Imperial College London, Blackett
 Laboratory, Prince Consort Road, London SW7 2AZ, UK}

\begin{abstract}
This paper analyzes a sample of 489 {\it Spitzer\/}/IRAC sources in
the Extended Groth Strip whose spectral energy distributions fit a
red power law from 3.6 to 8.0~\micron.  The median for sources with
known redshift is $\langle z\rangle =1.6$. Though all or nearly all
of the sample are likely to be active galactic nuclei, only 33\%
were detected in the EGS X-ray survey (AEGIS-X) using 200~ks {\it
  Chandra\/}\ observations.  The detected sources are X-ray
luminous with $L_X > 10^{43}$ erg~s$^{-1}$ and moderately to
heavily obscured with $N_H > 10^{22}$~cm$^{-2}$. Stacking the
X-ray-undetected sample members yields a statistically significant
X-ray signal, suggesting that they are on average more distant or
more obscured than sources with X-ray detections.  The ratio of
X-ray to mid-infrared fluxes suggests that a substantial fraction
of the sources undetected in X-rays are obscured at the
Compton-thick level, in contrast to the X-ray-detected sources, all
of which appear to be Compton-thin. For the X-ray-detected
power-law sources with redshifts, an X-ray luminosity $L_X \sim
10^{44}$~erg~s$^{-1}$ marks a transition between low-luminosity,
blue sources dominated by the host galaxy to high-luminosity, red
power-law sources dominated by nuclear activity.  X-ray-to-optical
ratios, infrared variability, and 24~\micron\ properties of the
sample are consistent with the identification of infrared power-law
sources as active nuclei, but a rough estimate is that only 22\% of
AGNs are selected by the power law criteria. Comparison of the
power-law selection technique and various IRAC color criteria for
identifying AGNs confirms that high-redshift samples selected via
simple IRAC colors may be heavily contaminated by
starlight-dominated objects.

\end{abstract}


\maketitle

\section{INTRODUCTION}

A  census of active galactic nuclei (AGNs) is a necessary
prerequisite for constraining
the accretion history of the Universe, describing the
origin of the cosmic background, and understanding galaxy formation.
X-ray detection is often used as an efficient method of finding
relatively uncontaminated samples of AGNs \citep{mus04}.  Large
numbers of AGNs can be found in deep X-ray surveys, where the AGN
density is an order of magnitude higher than that found in deep
optical observations via broad emission lines \citep[e.g.,][]{hal00, ste02, bau04}.  
X-ray surveys have now resolved $\sim$90\% of the
cosmic X-ray background (CXRB) at 2--8~keV \citep[e.g.,][]{mus00,
bau04, hic06}.
However, current X-ray telescopes lose sensitivity toward higher
energies.  Only 60\% of the X-ray background is resolved at 6--8~keV,
and this fraction drops further to 50\% at $>$8~keV 
\citep{wor04, wor05}.  Synthesis models of the hard X-ray background
spectrum require larger numbers of AGNs than are currently observed.
Heavily obscured, Compton-thick AGNs are expected to outnumber
unobscured AGNs by a ratio of $\sim$4:1 \citep[e.g.,][]{mai95, com01,
gil01, gil07}, though the exact value is still not well constrained.
Deep observations can sometimes detect the most luminous of these hard
X-ray sources \citep{toz06, geo07}, but even the deepest X-ray surveys
can miss large populations of obscured AGNs.

With the launch of highly sensitive infrared (IR) instruments aboard
the {\it Spitzer Space Telescope\/}\ \citep{wer04}, IR surveys have become a 
useful complement to X-ray and visible-light AGN searches.  AGNs
are thought to have a dusty environment (often imagined as a torus)
surrounding an optically- and X-ray-bright accretion disk centered on
a massive black hole (e.g., \citealt{Antonucci1993,Urry1995}).  Incident X-rays from the central
engine can be absorbed and reprocessed by the circumnuclear dust and
re-emitted at longer wavelengths, making AGNs IR-bright \citep{bar87, gra94,
nen02}.  The infrared bands are also much less affected by dust
extinction than the optical, making them ideal for identifying obscured
sources.

There have been several studies of AGN selection using {\it Spitzer\/}\ mid-IR colors
\citep[e.g.,][]{lac04, ste05, hat05} or other IR properties
\citep[e.g.,][]{mar05, don05, dad07, fio08,par08}.  One
straightforward way to search for obscured AGNs is to study their
spectral energy distributions (SEDs) \citep{alo06, bar06, don07,
car08, don08}.  In the mid-IR, AGNs can often be distinguished by
their red power-law SEDs, which can extend from the UV
all the way into the far-IR \citep{ree69, neu79,war87, elv94,Assef2010}.  This
behavior is thought to arise from a combination of thermal and
non-thermal components near the dusty nucleus that together produce a
broad  SED rather than from a single power-law emission source
\citep{rie81,bar87,Kotilainen1992,Klaas2001}.
Analysis of broadband SEDs therefore proves to be a powerful diagnostic,
differentiating AGNs from sources dominated by starlight. The latter generally
have a blue mid-IR power-law SED corresponding to the
Rayleigh-Jeans tail of the blackbody spectrum or else do not fit a
power law at all due to their 8~\micron\ PAH emission or 1.6~\micron\
stellar bump.

Finding large samples of AGNs in multiwavelength surveys requires both
wide area and good sensitivity.  This paper uses deep
{\it Spitzer Space Telescope\/}\ data to select a sample of 489 red
power-law sources in the Extended Groth Strip (EGS).  The EGS ($\alpha
= 14^h17^m$, $\delta = +52\arcdeg30\arcmin$) is a premier
extragalactic survey field observable from the Northern Hemisphere
that fits both criteria \citep{dav07}.  The EGS has been surveyed at
nearly every available wavelength to considerable depth and area, and it
spans more than $\sim$2\arcdeg\ by 10\arcmin\ in the 9.1~ks IRAC
pointings \citep{bar08}. The region's high ecliptic and Galactic
latitudes make it 
well-suited for studying extragalactic sources with little Galactic
contamination.

The unique combination of depth and size makes the EGS an ideal field
for an AGN search: it is  shallower but larger than the combined GOODS fields and
smaller but deeper than the COSMOS field at most wavelengths.  
Previous IRAC power-law studies have been done in fields
with much deeper X-ray exposures ($\sim$1~Ms) but shallower IRAC
observations \citep[$\sim$500~s, e.g.,][]{don07}.  The size and
depth of the EGS field in the IR bands allows us to probe different
redshift regimes with larger samples and improved statistics.  This
study therefore provides a view of power-law sources that is
complementary to previous work and extends the work of
\citet{bar06}, who analyzed the IR properties of
X-ray selected AGNs in a small portion of the EGS.  Multiwavelength
data are utilized to further study the properties of these AGN
candidates, including deep \emph{Chandra X-ray Observatory} pointings
to confirm AGN status and control contamination.

This paper proceeds as follows: \S2 introduces the data sets used,
and  \S 3 describes the power-law AGN selection process.  \S4
compares the power-law selection with other IR selection methods.
 \S5 discusses the sample redshift distribution and \S6 the  spectral energy
 distributions.  \S7 details the X-ray, IR,
 visible, and radio properties of the power law sample.    A summary is given in \S8.
 Throughout this paper, we adopt $\Lambda$CDM cosmology with $H_0 =
 71$~km~s$^{-1}$~Mpc$^{-1}$, $\Omega_M = 0.27$, and $\Omega_{\Lambda}
 = 0.73$.  All magnitudes are in AB units unless
 otherwise noted.

\section{OBSERVATIONS AND DATA REDUCTION} 

Infrared Array Camera \citep[IRAC;][]{faz04} observations of the EGS
were obtained in 2003--2004 as part of Guaranteed Time Observation
(GTO) program 8 with time contributed by G.\ Fazio, G.\ Rieke, and
E.\ Wright.  Each position was observed for approximately 9.1~ks at
all four IRAC wavelengths.  The data were processed by the {\it
  Spitzer\/}\ Science Center IRAC pipeline and custom software, and
sources were identified using SExtractor.  \citet{bar08} give details
of the observations and reductions and list source photometry in a
variety of apertures. This paper uses the photometry in the 3\farcs06 diameter
aperture, for which the data yield 50\% point-source completeness
limits of 1.1, 1.2, 6.3, and 6.9~$\mu$Jy at 3.6, 4.5, 5.8, and
8.0~\micron, respectively.\footnote{This paper actually used a
  not-quite-final version of the published catalog.  The only effect
  of this is that IRAC flux densities used here and given in Table~2
  are 3\% higher than the final values.  This has no effect on
  sample selection and trivial effect on ratios of IRAC to
  other-wavelength fluxes.} \citeauthor{bar08} also discussed the
observational uncertainties, which are used here with an additional
5\% of the flux density added in quadrature to allow for systematic errors.

The IRAC data are supplemented by observations from the Far-Infrared
Deep Extragalactic Legacy (FIDEL)
project\footnote{http://ssc.spitzer.caltech.edu/legacy/fidelhistory.html;
  PI: Mark Dickinson} taken with the Multiband Imaging Photometer
\citep[MIPS;][]{rie04} aboard {\it Spitzer\/}.  The FIDEL data fully
overlap the IRAC EGS pointings and include a $\sim$12~ks exposure at
24~\micron\ and a $\sim$5.4~ks exposure at 70~\micron.  Observations
used here are from the preliminary FIDEL catalog (dated 2008 Aug 28
--- M.\ Dickinson, private communication 2008).  Source extractions
were done with {\sc daophot}, which uses PSF fitting to determine
magnitudes.  The 70~\micron\ source catalog is 99\% complete for flux
densities above 5~mJy.  The limiting flux density at 24~\micron\ is
30~$\mu$Jy \citep{Salim2009}.  For the FIDEL data, our analysis uses
only the observational uncertainties calculated by {\sc daophot}.

The X-ray data were obtained with eight 200~ks pointings  using
the ACIS-I instrument on the {\it Chandra X-ray Observatory\/} and reach
a limiting flux of $5.3 \times 10^{-17}$~erg~cm$^{-2}$~s$^{-1}$\ ($3.8 \times
10^{-16}$~erg~cm$^{-2}$~s$^{-1}$) in the 0.5--2 (2--10)~keV band.  These
observations currently represent the third deepest X-ray survey in the
sky, and an extra 1.8~Ms of data have been taken in {\it Chandra\/}\ cycle
9 to observe the central fields of the EGS to a depth of 800~ks each.
The 200~ks data were reduced using the CIAO data analysis software
version~3.3 as described by \citet{lai08}.  Of the 1325 sources in the
existing AEGIS-X catalog, 882 are within the IRAC region.  At the
200~ks X-ray
depth, the majority of the detected sources should be active nuclei
based on their redshifts and luminosities: of the $\sim$250 cataloged
sources with IRAC counterparts and reliable redshifts, 95\% have
luminosities $L_{2-10~\rm keV} > 10^{42}$~erg~s$^{-1}$\ as expected for AGNs,
and all sources have $L_{2-10~\rm keV} > 10^{40}$~erg~s$^{-1}$.  In the EGS
field, a previous study \citep{bar06} found a stellar contamination rate of
$\sim$3\% among the X-ray identified objects.

Additional data used include visible photometry from the Suprime
camera on the Subaru telescope (S.~Miyazaki 2005, private
communication) with a 5$\sigma$ limiting AB magnitude of
$R \sim$27.  Additional visible-light data were taken in the $B$, $R$, and
$I$ bands from the Canada-France-Hawaii Telescope with limiting AB
magnitudes of 24.5, 24.2, and 23.5 respectively at 8$\sigma$
\citep{coi04} as well as in the $u'$ and $g'$ bands with Megacam
\citep{mcl06} at the MMT with 5$\sigma$ limiting AB magnitudes of 26.5
and 27.2, respectively (Ashby et~al.\ in prep.).  We used near-IR
observations in the $J$ and $K$ bands from the Wide-Field Infrared Camera
on the Palomar 5-m telescope \citep{bun06} with limiting magnitudes
of 23~mag (Vega) and 20.6~mag (Vega) at $5 \sigma$ for $J$ and $K$,
respectively.  All these visible and near-IR data sets span the entire
IRAC field except that the $J$-band images cover only one third of
the field.  Optical morphology was studied using $I$ (F814W) band
images from the Advanced Camera for Surveys (ACS) aboard the Hubble
Space Telescope (HST) \citep{lot08}.  We utilize 
the AEGIS20 radio survey \citep{ivi07}
which comprises 108\,hr of observations with the Very Large Array in
its B configuration and covers two thirds of the IRAC field with a
source sensitivity limit of 50~$\mu$Jy.  Finally, we
use spectroscopic redshifts obtained across two-thirds of the field by
the Keck DEIMOS spectrometer as part of the DEEP2 survey
\citep{dav03} as well as targeted redshifts available over the
whole field from 2006--2007 observations on the MMT Hectospec fiber
spectrograph \citep{Coil2009}.

The final parent sample for power-law selection is restricted to sources detected
in all four IRAC channels with signal-to-noise ratios of at least
five.  Because the X-ray data provide an important and straightforward
way of confirming AGN status, the analysis is further limited to IRAC
sources within the {\it Chandra\/}-observed field.\footnote{Restricting to
  the {\it Chandra\/}\ area eliminates only 5\% of the four-band IRAC
  sources.}
The resulting 0.35~deg$^2$ survey area contains
over 11,000 four-band IRAC sources as well
as over 700 X-ray sources with Poisson false probabilities $<\! 4
\times 10^{-6}$.

The multiwavelength datasets were cross-identified
by matching all sources to the IRAC catalog positions using a
1\farcs5 search radius.
This procedure nearly always resulted in a unique match, and in the
few ambiguous cases, the source closest to the IRAC position was
taken as the counterpart.  Comparing this approach 
to the likelihood-matching method used for the X-ray data  by
\citet{lai08} results in a discrepancy of less than 1\% in our final
AGN candidate sample.
Because {\it Chandra\/}'s point spread function (PSF) worsens with increasing
off-axis angles (OAAs),  in matching the IRAC sources to the X-ray
catalog, we checked for sources at large OAAs where the possibility of
a mismatch due to positional inaccuracies is higher.  The
{\it Chandra\/}\ pointing is wider than the IRAC field, and therefore 
many of the sources
with the largest OAAs do not lie within the shared area.  Out of 719
X-ray counterparts, only ten X-ray matches have OAAs greater than
9\arcmin, and four of these have OAAs greater than 10\arcmin.
Therefore, we are confident of the vast majority of the
IRAC-{\it Chandra\/}\ associations.

\section{POWER-LAW SAMPLE SELECTION}

In the infrared, the slope of a galaxy SED may be characterized by a
power law behavior of flux density with frequency $S_{\nu} \propto \nu^{\alpha}$.
AGN SEDs often follow a negative-sloping (red) power law, which may
arise from either thermal or non-thermal emission
originating near the 
central region \citep[e.g.,][]{ree69, neu79, elv94}. In contrast,
stellar-dominated sources at low redshift generally exhibit positive (blue) IRAC
power-law emission following the Rayleigh-Jeans tail of the blackbody
spectrum.  If $z\ga1$,  the SED of stellar-dominated sources may deviate from a power
law because of the 1.6~\micron\ stellar 
emission bump, which arises from the minimum in the opacity of the H$^-$
ion present in stellar atmospheres, but the overall slope is likely
to be blue or at least flat unless $z\ga2$.

For this paper, we defined a sample of
IR-selected  AGNs by choosing sources with SEDs that follow a
power law in the four IRAC bands from 3.6 to 8.0~\micron\ (observed,
not rest) wavelengths. 
A good fit was defined by small chi-square such that $P(\chi^2) \ge 0.1$, i.e., a
source having a true power law SED has a 10\% chance of being
excluded from the sample by random observational errors.
Galaxies fitting a power-law SED were further classified as red or
blue.  A limit of $\alpha \le -0.5$ was chosen to select our final
sample of AGNs  following \citet{alo06}, and throughout this paper, we refer
to ``red'' power laws as those with $\alpha \le -0.5$, while ``blue''
power laws refer to those with $\alpha > -0.5$. The cut-off was
based on the average visible--IR spectral slope of AGNs $\alpha \sim -1$
\citep{elv94, neu79} as well as the visible spectral indices of
$\alpha = -0.5$ to $-2$ of QSOs from the Sloan Digital Sky Survey
\citep[SDSS;][]{ive02}.  \citet{ste05}  found a similar slope in
the IRAC bands of $\alpha = -1.07 \pm 0.53$ among broad line AGNs.

\begin{figure}
\epsscale{0.5}
\plotone{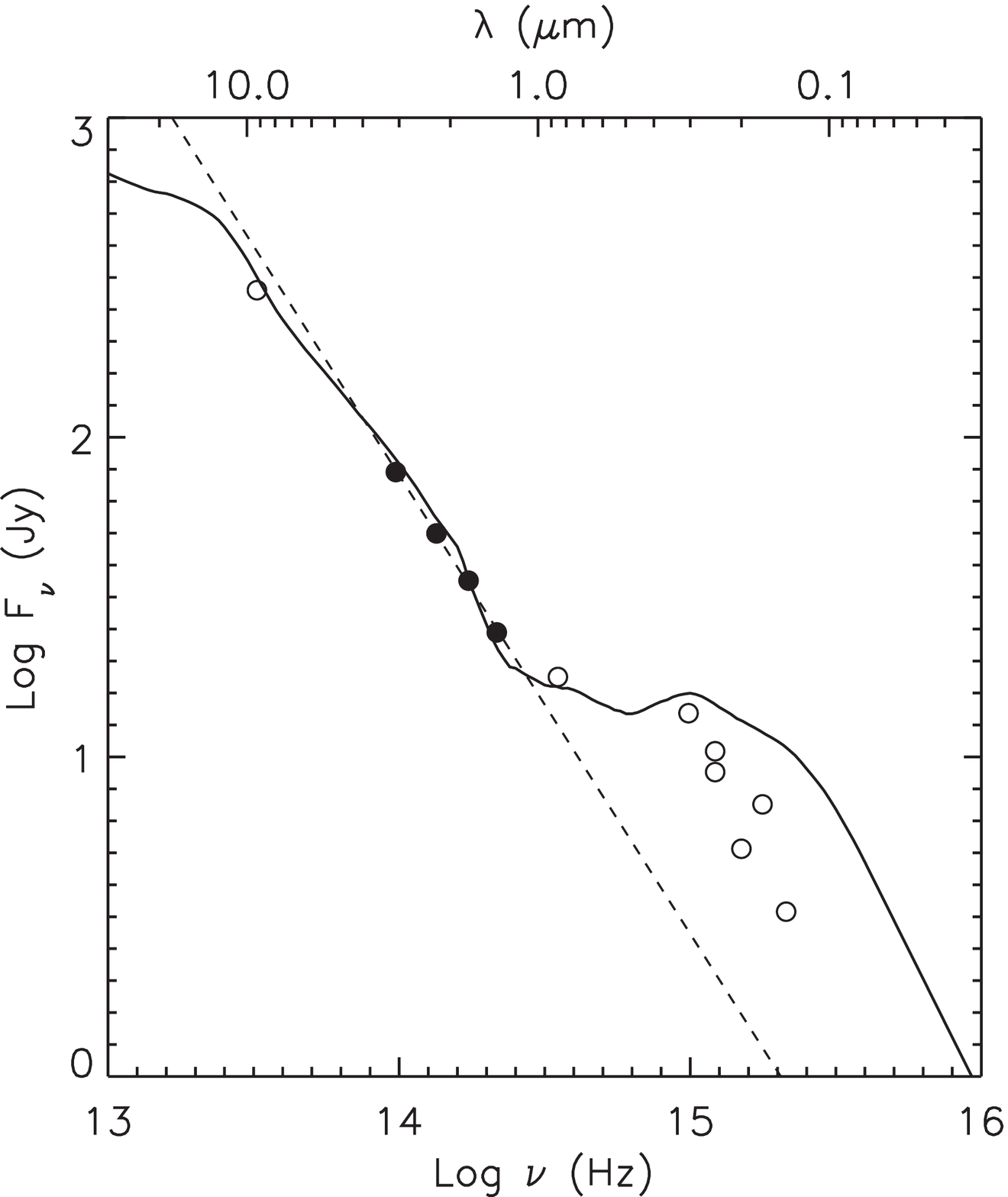}
\caption{Sample  SED and power-law fit for a
  PL AGN (EGSPL~81; see Table~\ref{tbl-2}).  The source
  was selected for having a spectroscopic $z=1.60$, near the
  median redshift 
  of the PL AGN sample.  Observed flux densities (not K-corrected) are plotted
  from $u$ band ($\sim$0.4~\micron) to 24~\micron\ with filled
  symbols denoting the four IRAC bands.  The wavelength scale is in
  the rest frame.  Uncertainties in IRAC flux
  densities are smaller than the  symbols.  The dashed line
  represents the fitted power-law line with spectral index $\alpha_i =
  -1.43$.  The solid line shows the median QSO SED from \citet{elv94},
  normalized at observed 4.5~\micron.  }
\label{plfitfig}
\end{figure}

\begin{figure}
\plotone{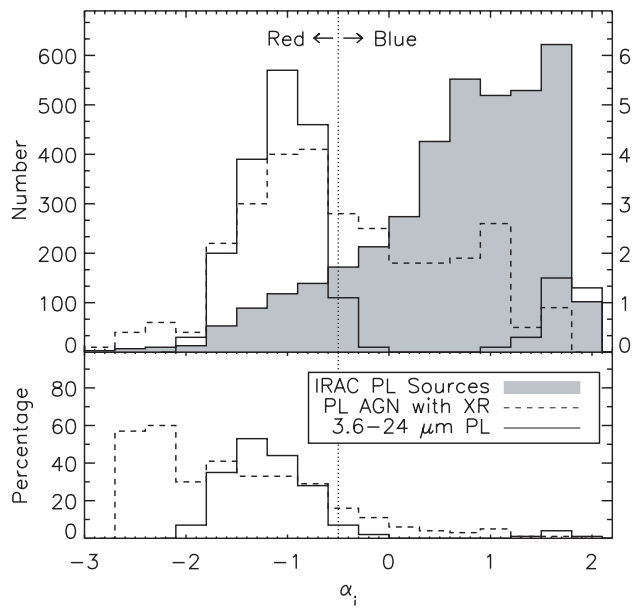}
\caption{Histogram of IRAC power-law spectral indices.
In the upper panel, the solid gray histogram represents all IRAC
sources that are well fit by a power law (red or blue); their numbers
are represented by the scale on the left.  The empty dashed
histogram denotes IRAC power-law sources with X-ray counterparts, and
the empty solid histogram shows sources well fit to a power law
from 3.6 to 24~\micron.  These histograms have been scaled to facilitate
viewing, and their  scale is on the right.  Sources
left of the vertical dotted line are defined here as PL AGNs.
The lower panel shows the percentage of power law
sources with X-ray counterparts (dashed lines) and the percentage of
sources well-fit by a power law between 3.6 and 24~\micron\ (solid
lines) out of all of the red and blue IRAC power-law sources
(represented by gray in the upper panel) as a function of spectral
index.  }
\label{alphahistfig}
\end{figure}

There are 489 sources within the IRAC/{\it Chandra\/}\ shared field that are
well fit to a red power law with $\alpha_{i} \le -0.5$, where
$\alpha_{i}$ is the spectral index from 3.6 to 8.0~\micron.  (See
Table~\ref{tbl-1} for a listing of fit categories and 
Figure~\ref{plfitfig} for a sample fit.  Uncertainties in slope for
individual objects range between $\pm 0.08$ and $\pm 0.21$.)  These
sources comprise only 4\% of the IRAC sources  in the field.
In comparison, 30\% of the IRAC sources are well fit by a power law
with $\alpha_{i} > -0.5$.  \citet{bar06} similarly found that 40\% of
IRAC sources fit a power law\footnote{``Fit'' is defined in their paper by the
less stringent criterion $P(\chi^2) > 0.01$.} with 7\% having
$\alpha_{i} < 0$ and 33\% having $\alpha_{i} > 0$. (In practice, the
differing cutoff between ``red'' and ``blue'' makes little
difference; see Fig.~\ref{alphahistfig}.)  Red power-law sources and
blue power-law sources have similar median flux densities at
8~\micron\ but diverge toward shorter wavelengths.  At 3.6~\micron,
the median flux density of the blue sample is five times higher than
the median flux density of the red sample.  The remaining 66\% of
sources do not fit a power law according to our criterion: 10\% have
red slopes with best-fit $\alpha_{i} < -0.5$ but $P(\chi^2) < 0.1$,
and 55\% have blue slopes but $P(\chi^2) < 0.1$.  The 489 sources
identified in our sample as having red power-law (PL) fits are
referred to here as PL AGNs and are listed in Table~\ref{tbl-2}. These sources were selected on the
basis of their IRAC spectral slopes only and therefore should not be
biased by X-ray or visible-light AGN-like properties.  The assumption that
these sources are AGNs will be justified below.

We also selected a sample of sources fitting a red power law from
3.6--24~\micron, thereby doubling the baseline used to determine
spectral slope.  The vast majority of MIPS sources (nearly all of
which were detected by IRAC) cannot be fit to a power law, red or
blue.  While most of the IRAC sources are generally brighter toward
the blue, the MIPS sources have much redder mid-IR emission, and only
0.5\% are well fit by a blue power law while 16\% are blue but poorly
fit by a power law.  82\% of MIPS sources are red but poorly fit by a
power-law, and finally 2\% (168) of MIPS sources are well fit by a
red power law from 3.6 to 24~\micron.  The predominance of red
sources is expected from the lower sensitivity limit of the MIPS
instrument compared to IRAC: IRAC sources detected by MIPS are likely
to be red unless they have extremely bright IRAC flux densities.  The
168 MIPS sources that fit a red power law are hereafter referred to
as the 3.6--24~\micron\ power law sample to distinguish them from the
PL AGN sample that is the main focus of this study.  All but 15 of
these 168 sources are in the PL AGN sample.  Of these 15, 12 are red
in the IRAC bands but were excluded from the PL AGN sample because
they do not fit a power law well enough at 3.6--8~\micron.  The other
three fit a power law in the IRAC bands but a blue one with
$\alpha_{i} \approx -0.35$ for all three sources.  These three
sources barely made the cutoff for a good 3.6--24~\micron\ red
power-law fit with $P(\chi^2) \approx 0.11$ and the 3.6 to
24~\micron\ slope $\alpha_{m} \approx-0.65$.

The distributions of fitted spectral slopes are shown in
Figure~\ref{alphahistfig}.  As shown, PL AGNs make up only a small
fraction of the total number of IRAC sources; their median spectral
index is $\alpha_{i} = -1.03 \pm 0.03$.  There is no significant
statistical difference between this calculated median slope and the
slope of the \citet{elv94} median QSO SED, which is $\alpha = -1.13$
at IRAC wavelengths.
Narrow-line AGNs  are expected to have steeper slopes than
broad-line AGNs, as their visible and near-IR emission is 
much more heavily obscured than their mid-IR emission, but there is
no evidence of  bimodality in the histogram of  PL AGN slopes.

\begin{figure}
\epsscale{1.1}
\plottwo{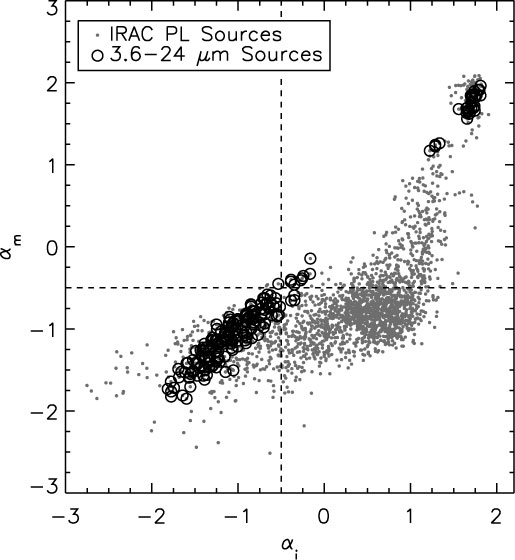}{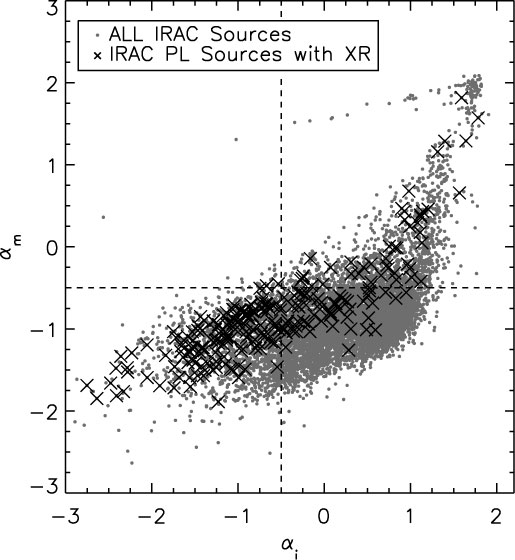}
\caption{Power law spectral slopes from 3.6 to
  8.0~\micron\ ($\alpha_i$) versus slopes from 3.6 to 24~\micron\ %
  ($\alpha_m$) for sources detected by both IRAC and MIPS.  Redder
  slopes are towards the bottom left in both panels. In the left  
  panel, sources well fit by a power law from 3.6 to 8.0~\micron\ are
  plotted as small gray dots and sources that fit a power law
  from 3.6 to 24~\micron\ as large black circles.  For comparison, the
  right panel shows the spectral indices of all  sources, including
  those poorly fit to a power law, as small gray dots.  Well-fit
  power-law sources with X-ray counterparts are overplotted with black
  X's.  The dashed lines in both panels
  indicate the spectral slope cutoffs for our AGN sample.  Objects
  with good fits plotted in the top right near $(1.8,2)$ are Galactic
  stars.  
     }
\label{alphacompfig}
\end{figure}

The 3.6--24~\micron\ power law objects do appear
to fall into two clear groups.  The majority of them are red
throughout the mid-IR
with $\alpha_{i} \le -0.5$; these have a median spectral
index of $\alpha_{i} = -1.11 \pm 0.03$.  There is also a small group
of sources that fit a blue power law from 3.6 to 24~\micron\ and
cluster at $\alpha_{i} \approx 1.8$, which is the slope 
of sources dominated by starlight.  These have colors 
$([3.6]-[8.0])_{\rm Vega} \approx 0$, as expected for
stars. 
For most sources, there is good agreement in spectral slope between sources
well fit by a power law in both 3.6--8.0~\micron\ and
3.6--24~\micron\ (see Figure~\ref{alphacompfig}). 
The X-ray sources span a wide range of $\alpha_{i}$ but
are overwhelmingly red in their 3.6--24~\micron\ colors.  These will
be discussed further in \S\ref{xray}.

The PL AGN with the steepest slope has $\alpha_{i} = -2.76 \pm 0.14$.
This is consistent with the maximum steepness found by other studies
and suggests a cutoff of mid-IR SED shape: \citet{alo06} found a
maximally steep visible-to-IR slope of $-2.8$ among their power-law
sources, and \citet{don07} found $\alpha = -3.15$.  The optically faint
X-ray sources in the CDFS studied by \citet{rig05} had a maximum
steepness of $\alpha = -2.9$, and \citet{sti96} found a limit of
$\alpha \sim -2.5$ among a sample of flat-spectrum radio sources.  
Figure~\ref{alphahistfig} shows that the PL AGNs with
the steepest spectral indices  never fit  a power law from
3.6--24~\micron.  Instead the SED turns over somewhere between 8 and
24~\micron, and the  actual 24~\micron\ flux densities of these
sources are several times lower than predicted by a
simple extension of the IRAC power law.

\section{PL SELECTION AND OTHER IR TECHNIQUES}

\subsection{Color-Color Plots}\label{color}

AGNs are generally redder than normal galaxies in the infrared,
and many studies have used this fact to select AGNs via their
positions in color-color space.    For example, \citet{lac04} and \citet{hat05}
defined separate AGN selections by studying luminous type 1 Sloan
Digital Sky Survey QSOs in the {\it Spitzer\/}\ First Look Survey and
the {\it Spitzer\/}\ Wide-Area Infrared Extragalactic Survey,
respectively, and \citet{ste05} plotted spectroscopically confirmed
AGNs in the Bo\"{o}tes\ field to define their selection region.

\begin{figure}
\epsscale{1.1}
\plottwo{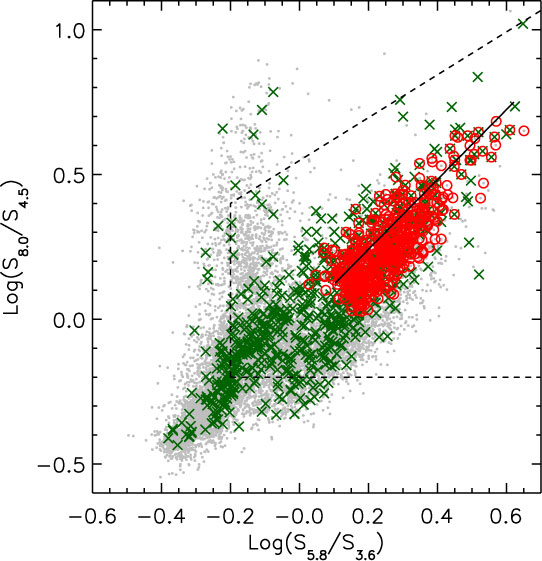}{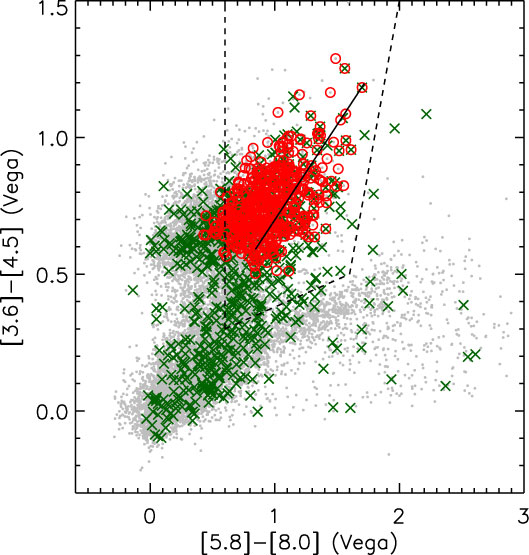}
\caption{IRAC color-color diagrams.  Dashed lines
  indicated the boundaries of the AGN selection wedges from
  \citet{lac04} (left) and \citet{ste05} (right).  Small gray dots
  represent the IRAC parent population, X-ray sources are denoted by
  green crosses, and the PL AGN sample is marked by red circles.  The
  solid diagonal lines indicate where power-law sources with spectral
  indices from $-0.5$ (lower left) to $-3$ (upper right) would lie.
  } 
\label{lacyfig}
\end{figure}

Power-law selection is similar in principle to the IRAC color-color
methods.  Both techniques search for luminous objects where
the nuclear activity outshines light from the host galaxy in the IRAC
bands, which sample the SED between the stellar emission  expected at
visible wavelengths and the dust emission at
mid-IR wavelengths.  However, the power-law selection is much
stricter, requiring in effect three color criteria rather than two
for selection of AGN
candidates.  For comparison, Figure~\ref{lacyfig} shows
our PL AGN sample  on the
color-color plots used by \citet{lac04} and \citet{ste05}. 
All of the PL AGNs are contained within the Lacy-defined
wedge.  However, the selection used by \citeauthor{lac04} is quite
broad; the wedge selects 60\% of the entire IRAC parent sample.  The wedge
also contains 80\% of the X-ray sources, but the X-ray sources
comprise only 8\% of the total sources in the AGN region. Even if the IRAC
catalog is truncated  to retain only sources that exceed the 
flux density limits of \citeauthor{lac04}, 40\% of the IRAC general
population is contained within the AGN region, though the X-ray
detection rate goes up to $\sim$40\%.  The \citet{ste05} selection is
less inclusive, containing just 18\% of the IRAC sample but 96\%
of our PL AGNs.
Fifty percent of the X-ray sources are contained by the wedge, and the
X-ray detection rate  within the wedge is 17\%.
The X-ray sources overall show a wide
range of colors and do not appear to lie within any well-contained
region of color-color space.  \citet{bar06} and \citet{car08}
similarly found that no IR color selection method can reliably select
all AGNs while avoiding the normal galaxy population.

Though color selections detect many low-luminosity AGNs due to the
broadness of the criteria, the depth of the IRAC data is important.
The EGS field, with its much deeper IRAC data, probes different
redshift regimes than the shallow fields studied by \citet{lac04},
who considered sources at $z \sim 0.3$, and by \citet{ste05}, whose
sample was mostly sensitive to galaxies at $z < 0.6$.  The results of
previous color-color AGN selection studies have been efficient and
reliable in part because they focused on low redshift regimes.  In
contrast, at the higher redshifts typical of our study, many non-AGNs
migrate into the AGN color selection wedge.  \citet{bar06} found that
while AGN SED templates stay within the AGN selection wedges at all
redshifts, starburst and normal galaxy templates have colors outside
the AGN wedges  at low redshift, but the colors migrate into the
AGN wedge at high redshifts. \citet{geo08a},
\citet{don08}, and \citet{Assef2010} similarly found that star-forming templates enter the
selection wedges both at low and high redshifts, though they avoid
the locus carved out by the power-law selection  (see Figure~6 of \citealt{bar06}
or Figure~4 of \citealt{don08}).  The templates
suggest that a high degree of stellar contamination is unavoidable
when the color selection technique is applied to deeper samples.  
Contamination can be reduced by restricting the sample to relatively
high 24~\micron\ flux densities, but \citet{don08} found that even
for $S_{24}>0.5$~mJy, half of the color-selected sources in their
sample cannot be confidently identified as AGNs.  
Power-law selection is a much more reliable method of finding AGNs:
most of the secure AGN identifications among the color-selected
galaxies in the GOODS-S field were contained in the power-law sample \citep{don08}.

\begin{figure}
\epsscale{.9}
\plotone{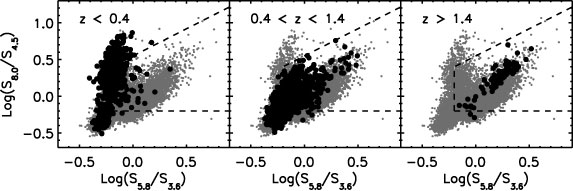}
\caption{IRAC color-color plots in different redshift
  bins.  The small gray dots represent the entire IRAC population
  regardless of whether they have redshift measurements, while
  the large black dots indicate the IRAC sources with spectroscopic
  redshift measurements that fall within the appropriate redshift
  bin. The dashed lines enclose the \citet{lac04} selection
  wedge. The plot includes all IRAC sources with spectroscopic
  redshifts, not   just PL AGNs. 
  }
\label{zevolfig}
\end{figure}

The expectation from templates is confirmed by the EGS IRAC
data. Figure~\ref{zevolfig} shows the evolution of IRAC colors for sources with
and without measured redshifts.  At $z < 0.4$, the 6.2~\micron\ and 7.7~\micron\ PAH
features can still dominate the 8.0~\micron-band emission, and the
plot is largely populated by sources with roughly constant blue
$S_{5.8}/S_{3.6}$ colors but spanning a wide range of red
$S_{8.0}/S_{4.5}$ colors.  These are likely to be nearby, stellar-dominated
galaxies with strong aromatic emission redshifting through the
8~\micron\ band.  At intermediate redshifts $0.4 < z < 1.4$, sources
migrate closer to the AGN  region of the color-color plot
as the K-correction shifts objects towards the red.  At $z >
1.4$, nearly every object has colors consistent with the AGN
wedge.  Thus, though we cannot assess starburst contamination at
high redshifts due to the biases inherent in our redshift survey (see
\S\ref{redshift}), we
reiterate the need for caution  to avoid contamination from
star-forming galaxies in using IRAC color-color
selection of AGNs.  We also caution that the color changes with
redshift seen in  Figure~\ref{zevolfig} should not be interpreted as
population evolution because the IRAC bands sample different rest
wavelengths at each redshift.

\subsection{Visible to IR Colors}

Another technique for finding AGNs is to select on very red
infrared-to-visible or infrared-to-UV color \citep[e.g.,][]{dad07, dey08, fio08,
  pol08,Pope2008}. This technique has principally been applied to
find X-ray-faint (possibly Compton-thick) AGNs. Adopting the approach
of \citeauthor{fio08} selects 89 sources in the EGS with
$S_{24}/S_{R} \ge 1000$, $R-K > 4.5$, and $S_{24} >40$~$\mu$Jy.  Only
two of these fit a red IRAC power-law.  The nature of the
``IR-excess'' selection is to choose optically faint sources, $R>
25.5$~mag, and as a result only one of the 89 IR-excess sources in the EGS has a
spectroscopic redshift  (with $z=1.1$).  Though the
majority of these galaxies cannot be fit to either a red or blue
power-law, $\sim$75\% have blueward-sloping SEDs. \citet{dey08} found
a similar result for IR-excess galaxies, at least for $S_{24} \la
0.8$~mJy: most have SEDs containing a stellar emission
component. Thus at least a major portion of the IR light in
IR-excess-selected objects is likely to be starlight, even if an AGN
is also present.

\subsection{Contamination of the Power-law Method}\label{contamination}

Little contamination is expected from stellar-dominated sources in
our PL AGN sample.  
Systems with infrared emission
dominated by star formation and also galaxies with high levels
of starlight from an older stellar population show curvature in their
SEDs and thus will not enter the 
PL AGN sample.  As shown below, the multiwavelength properties of
the PL AGN sample also point to strong nuclear activity. The red IRAC
colors of the PL AGNs are unlikely to arise from PAH emission in the
8~\micron\ band; such emission would not produce
an overall power law SED. Furthermore, of the
sources with redshifts (\S\ref{redshift}), only four have $z < 0.5$,
and none has $z < 0.4$.  For higher redshifts that characterize the
bulk of the PL AGNs, the strong PAH features are outside the IRAC
bands.  While the PL AGNs may have stellar emission at some level ---
even enough to be a major component of the bolometric luminosity --- it
should not dominate the light observed at IRAC wavelengths.

PL AGN SEDs generally follow AGN templates with no evidence of stellar
bumps, but an attempt to model the UV-MIR SEDs of the PL AGNs more
precisely was inconclusive.  Unfortunately, PL AGNs SEDs offer little
information because of their lack of spectral features.
The possibility that some objects in the PL AGN sample are really
starbursts  cannot be directly
tested because, if they are starbursts, they must have photometric redshifts $z \ge 2$, while
spectroscopic measurements  at these high redshifts are limited to
AGNs (\S\ref{redshift}).  Modeling the UV--MIR SEDs of X-ray undetected PL
AGNs is particularly difficult because most of these sources have very low or
undetected visible flux densities.  However,  our X-ray
stacking analysis (\S\ref{stacking}) demonstrates that these sources have hard X-ray
emission and are likely to be obscured AGNs.

We also studied the SED templates for a luminous infrared galaxy
(LIRG) and an ultraluminous infrared galaxy (ULIRG) made from average
composite fits of observed LIRGs and ULIRGs \citep{don08}.  An IRAC
power-law shape cannot be obtained simply by redshifting the (U)LIRG
SED templates we tried to the maximum redshifts ($z\approx3.5$) that
we observe.  \citet{alo06} suggested that cool ULIRGs at $z > 2$ and
with $\alpha_{i} > -1$ could contaminate a power-law AGN sample, but
the SED templates in their Figure~11 show curvature resulting from
the 1.6~\micron\ stellar bump in the IRAC wavelength range.  The
stellar bump should exist whenever a substantial fraction of the rest
1.6~\micron\ light comes from evolved stars, and its absence (by
definition) in our PL AGN sample should eliminate most if not all
cool ULIRGs.  More recently, \citet{nar09} have presented theoretical
models in which reddened starlight from a young population plus an
AGN component can exhibit a power law SED during brief phases of a
merger.  There are 21 PL AGNs with spectroscopic $z>2$, but their
visible spectra all show AGN emission lines (\S\ref{redshift}).
Moreover, all but four have X-ray hard-band 
detections with $L_{2-10 \rm keV} > 10^{44}$~erg~s$^{-1}$.  Members
of our PL AGN class that are too faint in visible light to have
spectra could conceivably be high-redshift ULIRGs of the type
proposed by \citeauthor{nar09}, but the brevity of the power-law
phase suggests such contaminants should be uncommon.  The
multiwavelength data on the PL AGN sample also suggest little
contamination.  Redshifts and thus luminosities for any star-forming
galaxies contaminating the PL AGN sample would have to be very high.

\section{REDSHIFT DISTRIBUTION AND EVOLUTION}\label{redshift}

\begin{figure}
\includegraphics[trim=0 72 0 72, clip=t,width=\textwidth]{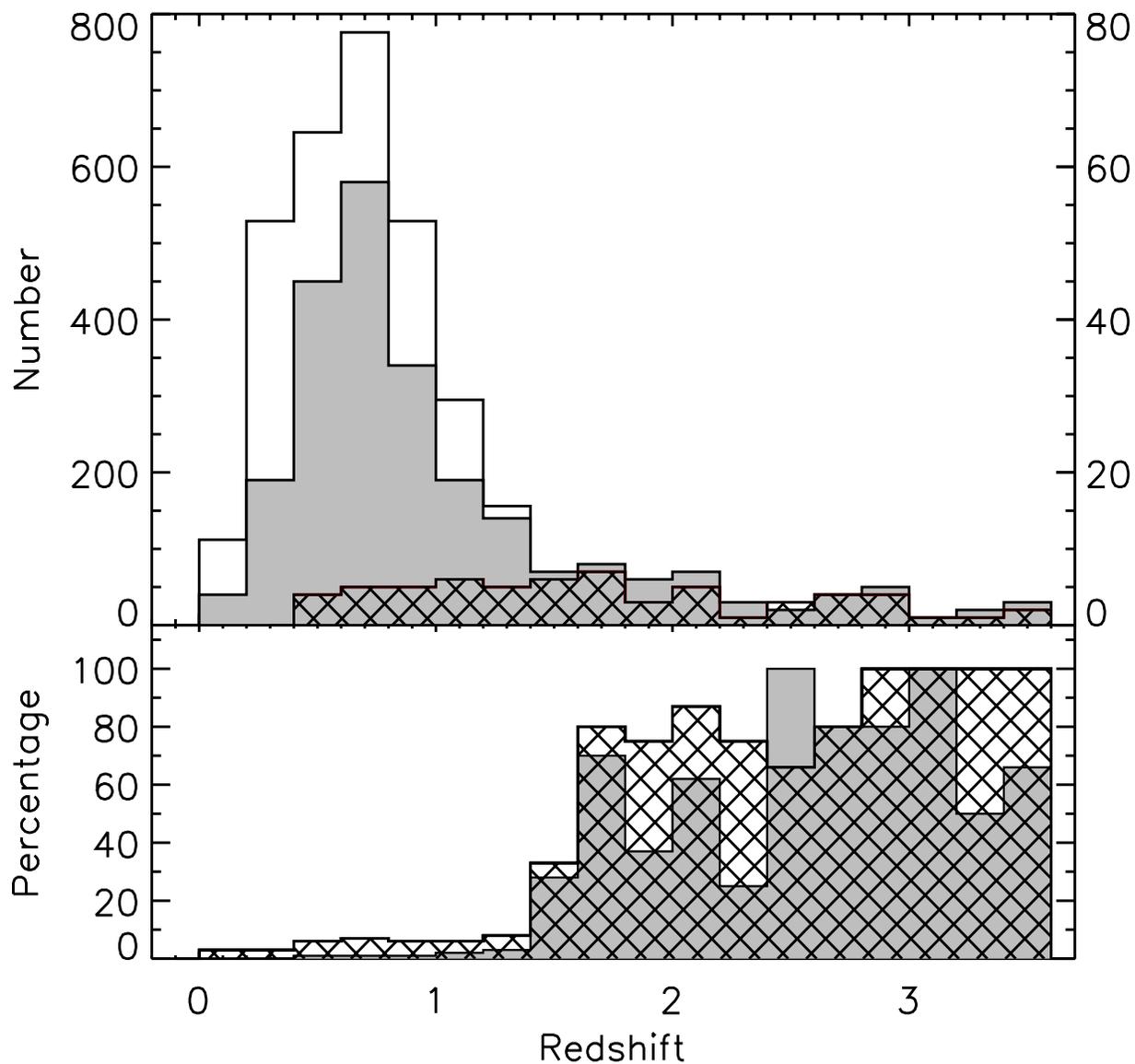}
\caption{
Histograms of spectroscopic redshifts. The empty
  histogram represents all IRAC sources with reliable redshifts (left
  ordinate), gray
  denotes sources with X-ray counterparts (right ordinate), and
  hatched lines denote 
  PL AGNs (right ordinate).    The bottom panel
  shows the percentage of X-ray (gray) and PL AGN (hatched)
  sources comprising the 
  known-redshift population (empty histogram in upper panel) as a function of
  redshift. Nearly all IRAC sources with measured redshifts $\ga$1.5
  are both X-ray sources and PL AGNs.}
\label{zhistfig}
\end{figure}

Because the power-law selection tends to choose sources that are
optically faint (\S\ref{optical}), the PL AGN sample has a smaller rate of redshift
measurement than the general IRAC sample.  This is true despite the
fact that many of the Hectospec redshifts \citep{Coil2009} specifically targeted red
power-law sources, X-ray sources, and radio galaxies.  Only 13\% of the PL AGNs have
spectroscopic redshifts compared to 26\% of the parent IRAC
population.  Part of the problem is that the comparatively deep
IRAC data probe higher redshifts than the typical redshift of the
spectroscopy or the X-ray data.  Another problem is that the
wavelength range of the DEEP2 observations was chosen to measure
galaxies with $z \la 1.4$, and at higher redshifts, typical galaxies
show no spectral features observable by DEEP2.\footnote{In
  particular, the $\lambda$3727\,\AA\ [\ion{O}{2}] doublet shifts out
  of the DEEP2 spectral coverage for $z > 1.4$.} 
Figure~\ref{zhistfig} shows that nearly all objects that have
spectroscopic redshifts $>$1.4 are PL AGNs.  Redshifts for these
sources come from ultraviolet lines that are prominent in AGNs but not
in normal galaxies.  Furthermore, to have been targeted at all for
redshift measurement, galaxies at such high redshifts must be
extremely luminous, and such high luminosities in the rest
ultraviolet (observed $R$ for DEEP2 selection) are most easily
provided by AGNs and moreover {\em unobscured} AGNs.  Indeed,
of 26 PL AGNs  with DEEP2 spectra,\footnote{quality 2 for three
  objects, $\ge3$ for the rest} 15 show strong lines and are classified ``AGN.''
The automated classifier called the remaining 11 ``Galaxy,'' but in
fact all 11 have emission lines, and at least four have broad
lines.  Only for three objects are broad lines clearly absent, and
the largest redshift for any of these is  $z=1.40$.
For the Hectospec redshifts, the target selection mentioned above forces most of the 
sample to be AGNs, and as expected
$>$90\% of this group was classified that way.  The few classified
``Galaxy'' are all at $z<1$.\footnote{The Hectospec sample includes one
  object at $z=2.18$ classified ``Galaxy'' by the automated
  classifier, but inspection (A.\ Coil 2010, private
  communication) shows it is really a broad line AGN, and the large
  redshift is correct.}
Taking the  strong selection effects into  
account, it is unsurprising that virtually all of the sources
observed to have  high redshifts are AGNs.

PL AGNs for which measured redshifts exist tend to have much higher
redshifts than the IRAC parent population.
The general population of IRAC sources in the DEEP2 redshift survey
has a median  $z = 0.72$, and the sources observed with
Hectospec have median $z = 0.51$.  The combined redshift sample has a
median $z = 0.68$.  Figure~\ref{zhistfig} shows the redshift distribution.
Sixty two of the 489 PL AGNs have high quality spectroscopic redshift
measurements, 39 from the Hectospec instrument at the MMT and 23
from the DEEP2 redshift survey using the Keck DEIMOS Spectrometer.
The median redshift of these sources is 1.6, and the maximum is 3.5.
The PL AGN redshifts are also significantly higher than the median $z = 0.8$
for IRAC sources with X-ray detections.  The redshift
distribution is relatively flat, suggesting that neither aromatic
emission features nor silicate absorption had much effect on sample
selection, in contrast to flux-limited  24~\micron\ samples
(e.g., \citealt{Huang2009}).  Of the 3.6--24~\micron\ power-law sources, 20 
have redshifts.  The median redshift for these sources is again high
at $z = 1.6$.

Though the selection effects explain why all the sources with
measured high redshifts are AGNs, it is interesting that so many of
the spectroscopic high-$z$ AGNs are identifiable as AGNs via their
mid-IR properties. In Figure~\ref{lacyfig}, for example, X-ray AGNs
exhibit a wide range of mid-IR colors, yet Figure~\ref{zevolfig}
shows that all the AGNs with high measured redshifts are
confined within a narrow color range. 
The reason is likely to be that at high redshift, AGNs
are seen and selected for spectroscopy only if they are luminous, and
in that case, the power law component is likely to dominate the SED.

\section{VISIBLE-TO-MID-INFRARED SEDS}\label{sed}

\begin{figure}
\epsscale{0.7}
\plotone{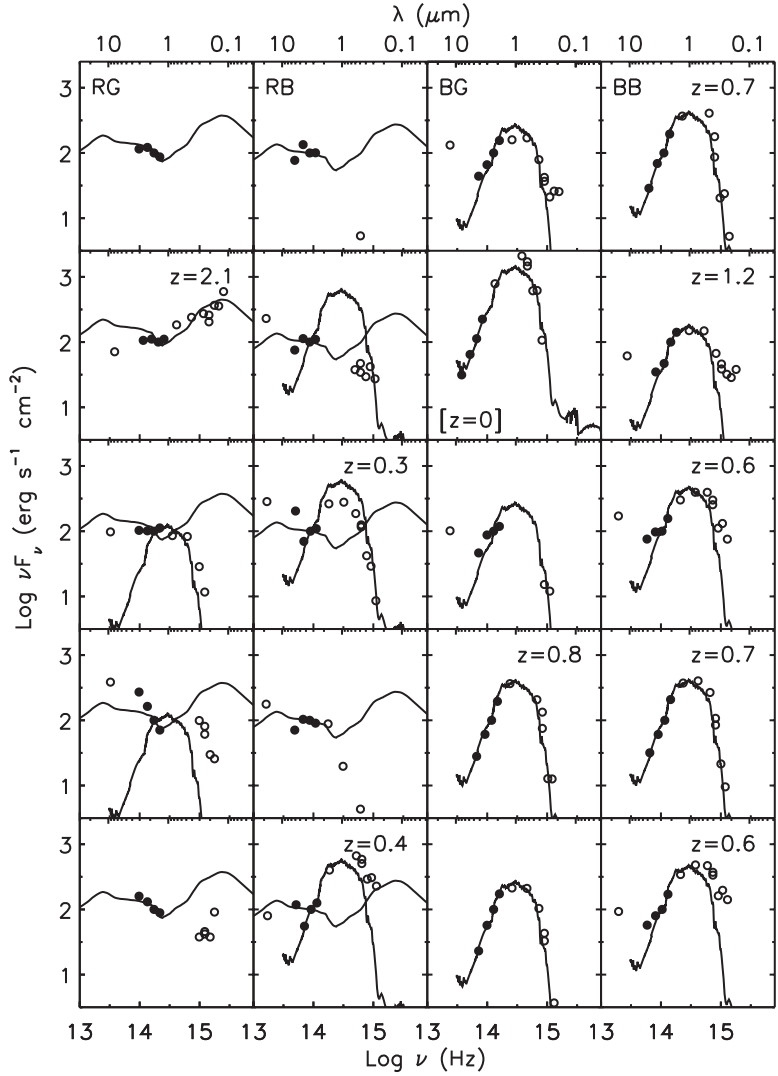}
\caption{\footnotesize
   SEDs from a random subset of sources in the general IRAC
  population.  The first column plots PL AGNs (RG = red,
  good fit: 4\% of IRAC sample); the second column plots IRAC sources
  with generally 
  red-sloping SEDs but which are poorly fit to a power law (RB = red,
  bad fit: 10\% of sample).  The third and fourth columns plot IRAC
  sources well fit and poorly fit respectively to a blue power law,
  (BG = blue, good fit: 30\% of sample; BB = blue, bad fit: 55\% of
  sample).  For 
  comparison, an elliptical template \citep{col80} is overplotted in
  the last two columns and five frames of the first two, and a median
  QSO SED \citep{elv94} is overplotted in the first two columns.
  Fluxes are in arbitrary units.    Data are plotted at
  estimated rest frame wavelengths; redshifts for individual objects
  are indicated where known.  Otherwise, data are plotted at the
  median redshift for the SED type in each column: 1.6 for RG, 0.30
  for RB, 0.92 for BG, and 0.70 for BB except for the second object
  in the BG group, for which data are plotted at $z=0$ even though no
  spectroscopic redshift is available. Templates are in the rest
  frame and are normalized to the observed 4.5~\micron\ flux density.
}
\label{allsedfig}
\end{figure}

The SEDs of PL AGNs generally do not exhibit the spectral behaviors
typical of stellar systems but tend to resemble Type~1
AGNs \citep[e.g.,][who fit template SEDs to IRAC/X-ray AGNs in the
EGS]{alm08}.  To compare the SED shapes,
Figure~\ref{allsedfig} shows the SEDs for a sample of randomly
selected sources from the general IRAC population in every category
of power-law fit.
There are 1156 sources that have red-sloping SEDs not fitting a power
law; these comprise 10\% of the total IRAC population
(Table~\ref{tbl-1}).  These sources are a diverse group,
some showing  a stellar bump and others not.  Compared to the other three
categories of sources, they have a higher rate of reliable redshift
measurement. (37\% have redshifts as opposed to only 13\% of the PL
AGNs.)  The median redshift of this group is $z = 0.3$, lower than that
of any other fit category.  Eighty-seven percent of these sources have
$z < 0.5$, and only 6\% have  $z > 1$.
These red non-PL objects are far more likely to be
detected at 24~\micron\ and 70~\micron\ and have much higher mid-IR
flux densities than the other sources.  They also have a higher rate
of radio flux detection, though their mean radio flux density is
\emph{lower} than that of the other three categories of fit.  They have a
higher rate of X-ray detection than the blueward sloping sources,
though the rate is not as high as seen among the PL AGN sample (see
\S \ref{xrrate}).
The majority of these sources are poorly fit to a red power-law
because their 4.5 and 5.8~\micron\ flux densities are lower than what
a power-law from 3.6$-$8.0~\micron\ would predict. A plausible
explanation for such SEDs is starlight emission, possibly heavily reddened, at 3.6~\micron,
thermal dust or AGN emission at 8~\micron\ (possibly including PAH
emission for low-redshift galaxies), and an SED minimum between these
wavelengths where neither component emits strongly.  Such an SED is
typical of nearby spiral galaxies \citep[e.g.,][]{Dale2005,Dale2007} or
galaxies with starlight contributing most of the light at 3.6~\micron\ but an
AGN dominating at 8~\micron\ \citep[e.g., NGC~3079 or 4826 ---][]{Lawrence1985}.  Thus this
group is likely comprised of a mix of relatively low-redshift, starlight-dominated galaxies 
and AGNs whose IR signatures are diluted or veiled by starlight.

All of the blue sources (85\% of the IRAC sample) show a broad
emission peak, typically reaching its maximum near 1--2~\micron. This
is an indication of stellar emission, as shown by the templates in
Figure~\ref{allsedfig}.  For the 27\% of the group with measured
redshifts, the mean redshift is 0.73; a range of 0 to 1.5 is
consistent with the wavelengths of SED peaks in the group.  While
some of these sources no doubt host an AGN, its emission is only a
minor component at 3.6~\micron.

\begin{figure}
\plotone{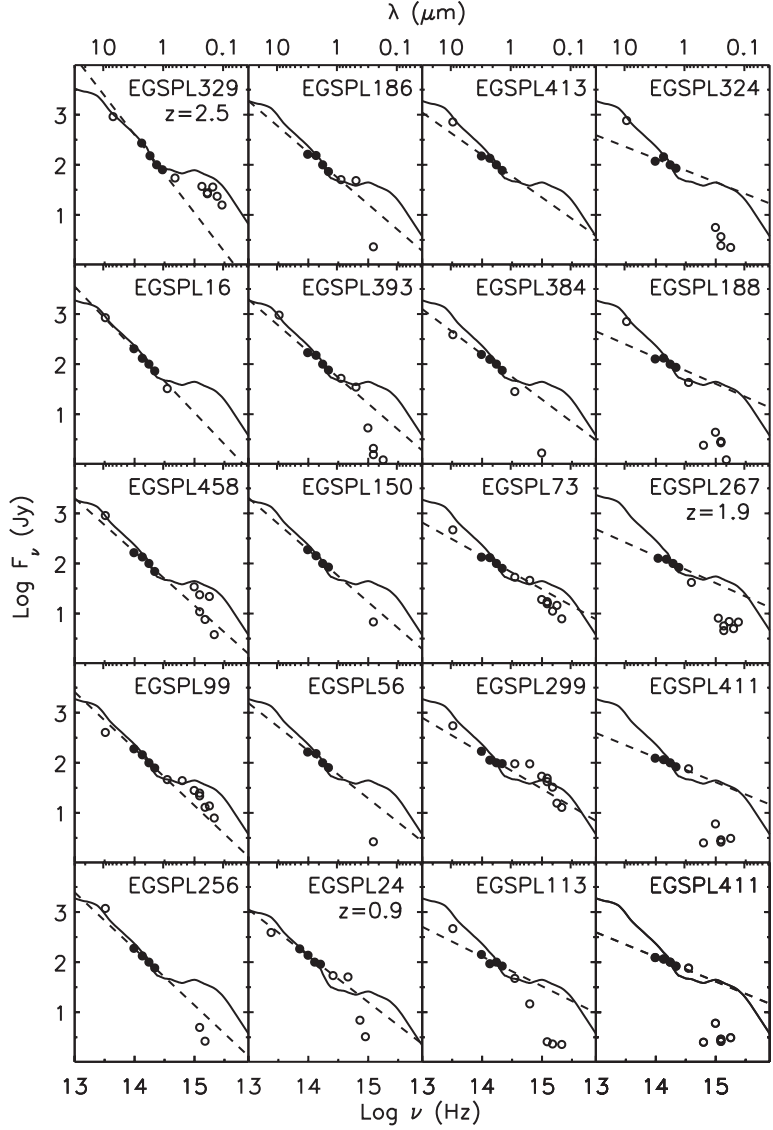}
\epsscale{.9}
\caption{\footnotesize
SEDs of a random subset of PL AGNs.  SEDs are
  sorted according to spectral index, with $\alpha = -1.6$ in the top
  left corner and $\alpha = -0.5$ in the bottom right corner.  Flux
  densities are plotted from $u$ band (0.4~\micron) to 24~\micron\ with
  arbitrary units. (See Table~\ref{tbl-2} for actual flux densities;  source identifications are indicated
  in the upper right of each plot.) Filled
  symbols represent the four IRAC bands.  The 
  dashed lines represent the best fit power law
  through the IRAC data points.  The solid lines show the
  \citet{elv94} median QSO template anchored at 4.5~\micron.   The
  sources are plotted at rest frame if a 
  redshift is listed below the ID.  If no redshift is listed, the
  source does not have a reliable redshift measurement and has been
  shifted to the median $z=1.6$.}
\label{plagnsedfig}
\end{figure}

Figure \ref{plagnsedfig} shows  SEDs for a selection of  PL AGNs.  As mentioned above,
the PL AGN  selection favors optically faint sources, and
18\% of the sample do not have $R$-band detections.  Of the sources
that are detected, 80\% have $R$-band flux densities that are more
than 50\% below the values predicted by the \citet{elv94} template,
while 9\% have 
flux densities more than 50\% above template values.  The X-ray
detection rate among the optically brighter sources is high (84\%),
while the detection among optically fainter sources is only 25\%.
These optically fainter sources have higher X-ray hardness
ratios\footnote{Hardness ratio ${\rm HR} = (H-S) / (H+S)$, where $H$ is the
2--7~keV count rate and $S$ is the 0.5--2~keV count rate.  See
\S\ref{hr} for further discussion of hardness ratios.} (${\rm HR} =
-0.1$ compared to ${\rm HR} = -0.4$ for the optically brighter sources),
suggestive of gas and dust extinction at visible and X-ray wavelengths.
Thus the power-law selection appears to be fully capable of finding
obscured AGNs.

\section{PL AGN MULTIWAVELENGTH PROPERTIES}\label{multi}

\subsection{X-Ray Properties}\label{xray}

\subsubsection{X-ray Detection Rate} \label{xrrate}

The power-law selection tends to find sources that have not been
identified in X-ray observations, even though PL AGNs have a
relatively high X-ray detection rate compared to the general
population.  Among the sample of 489 PL AGNs, 159 (33\%) have X-ray
counterparts in at least one band in the 200 ks AEGIS-X catalogs
\citep{lai08}, which require Poisson false probabilities $<\!4 \times
10^{-6}$.  (Table~3 summarizes the X-ray detection rates.)
The 3.6--24~\micron\ power-law sources have a slightly
higher rate of strong X-ray detection with 65 out of 168 (39\%)
sources having AEGIS-X counterparts.  The parent sample of IRAC
sources within the {\it Chandra\/} field has an X-ray detection rate
of only 6\%, and the IRAC sources with 24~\micron\ counterparts have
an X-ray detection rate of 7\%.  For comparison, in fields with much
deeper X-ray data of $\sim$1~Ms, \citet{alo06} found that 53\% of
their power-law galaxies in the CDFS were X-ray detected, while
\citet{don07} found an initial detection rate of 55\% with Poisson
false probabilities $< 1 \times 10^{-7}$ in the CDFN\null.  Thus
though deeper X-ray data find more PL AGNs, even the deepest X-ray
data available leave a substantial fraction of IR power-law galaxies
undetected.  More typical X-ray depths of a few hundred ks detect
only about half the PL AGNs.  Lowering the detection threshold
increases the detection fraction but still leaves substantial numbers
of PL AGNs undetected.  In the EGS, lowering the detection threshold
to false probability $<\!2 \times 10^{-2}$ (corresponding to
2$\sigma$ in Gaussian statistics) only raises the X-ray detection
rate of PL AGNs to 40\%.  In the deepest field, the CDFN, 85\% of PL
AGNs were detected at the 2.5$\sigma$ significance level \citep{don07}.

\begin{figure}
\epsscale{1.1}
\plottwo{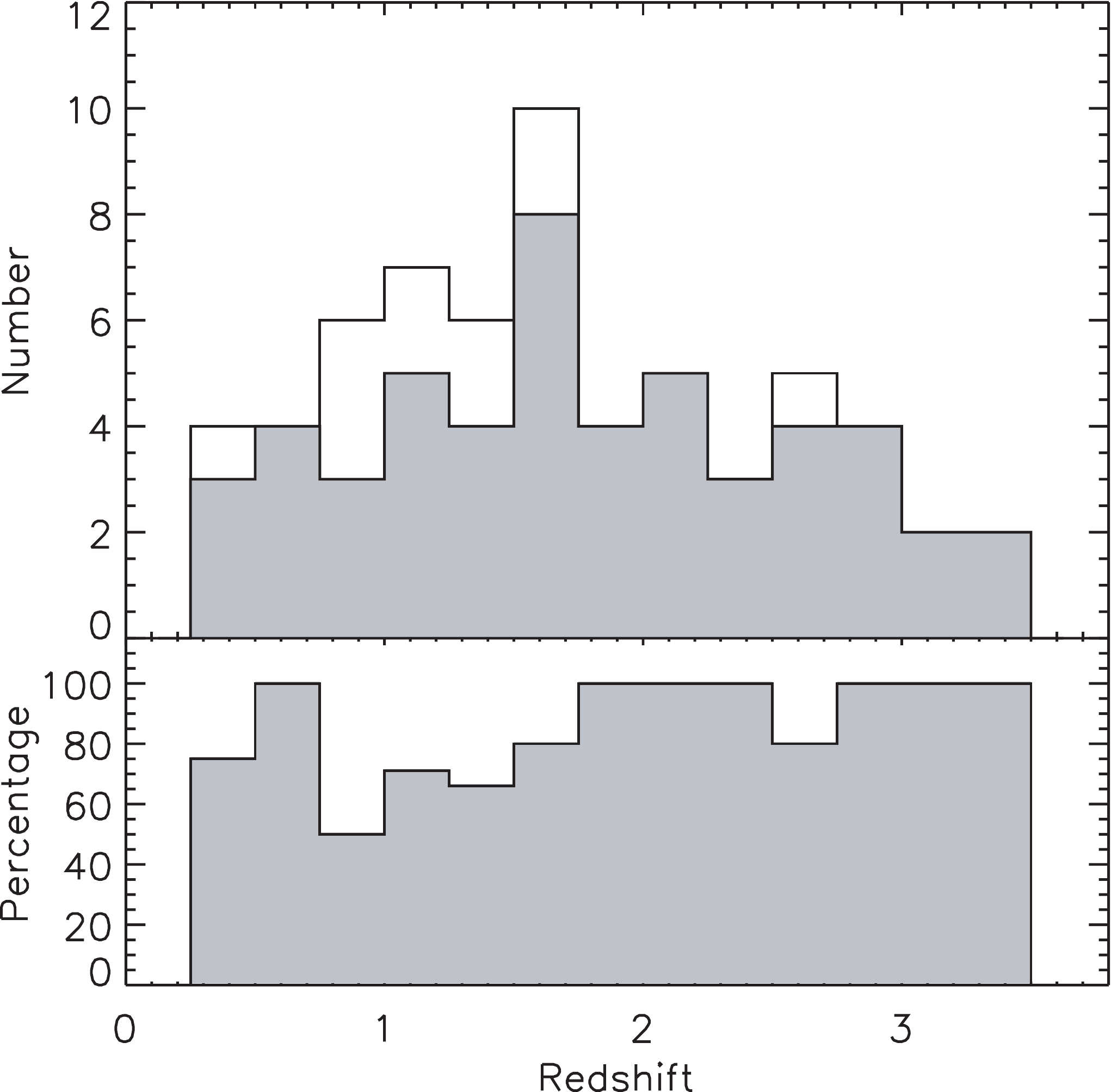}{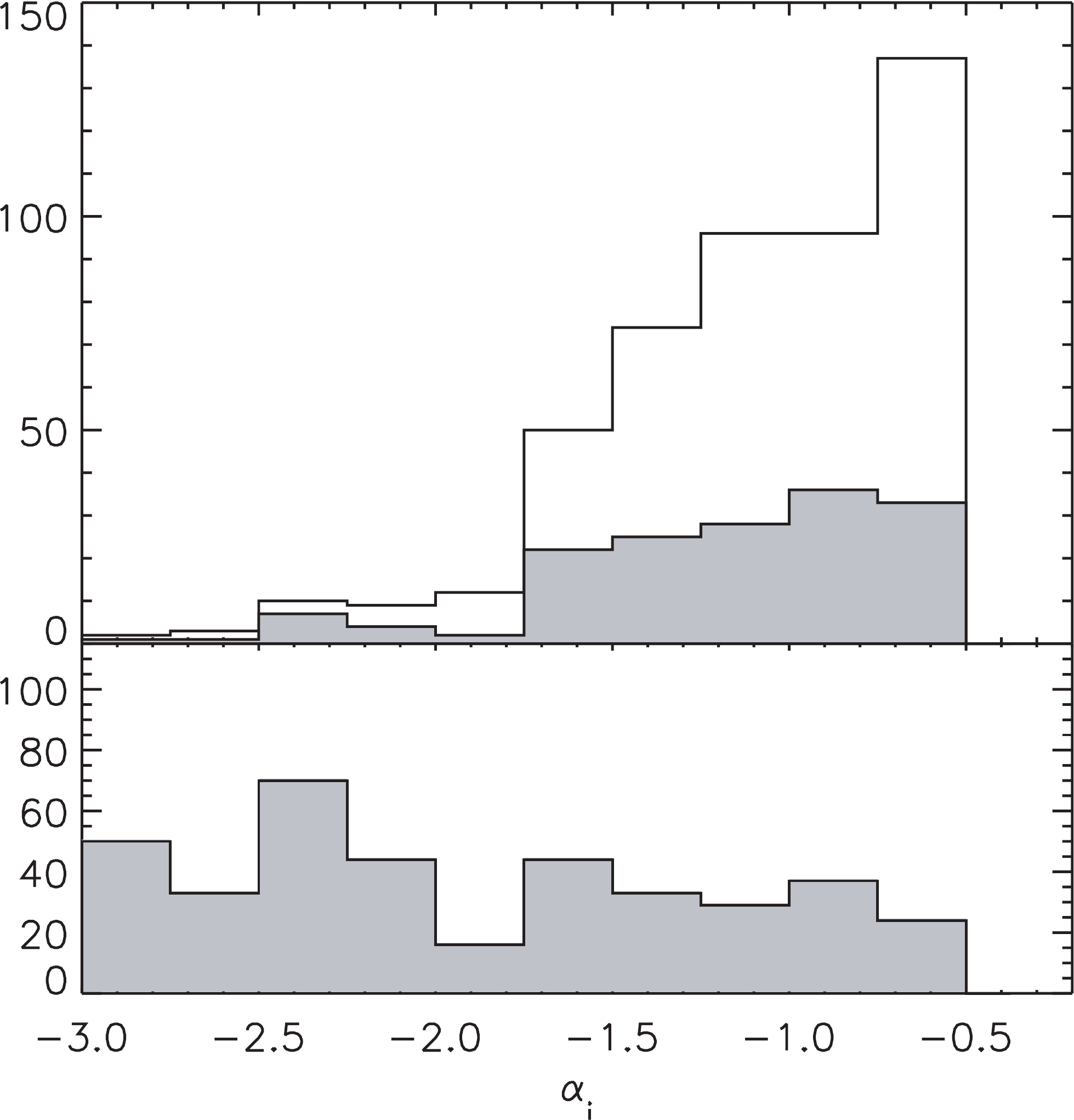}
\caption{X-ray detection fractions of PL AGNs as a function of
  spectroscopic redshift (left) and IRAC spectral index (right).  In
  the top panels, the empty histogram plots all PL AGNs, and the gray
  histograms denote PL AGNs with X-ray counterparts.  The bottom
  panels show the X-ray detection percentages among the PL AGN sample
  as a function of redshift and spectral index.}
\label{xpropfig}
\end{figure}

The fraction of X-ray-detected PL AGNs does not
vary significantly with IRAC spectral slope (see
Figure~\ref{xpropfig}).  The median IRAC fitted index among the X-ray detected
PL AGNs is $\alpha_{i} = -1.09 \pm 0.05$ compared to the median of
$\alpha_{i} = -1.00 \pm 0.03$ for PL AGNs without X-ray counterparts.
For the 3.6--24~\micron\ power-law sources, the median index is $\alpha_{m} =
-0.98 \pm 0.05$ for the X-ray detected sources and $\alpha_{m} = -1.15
\pm 0.04$ for X-ray non-detected objects.

\begin{figure}
\includegraphics[trim=0 0 0 144,clip=t,width=6in]{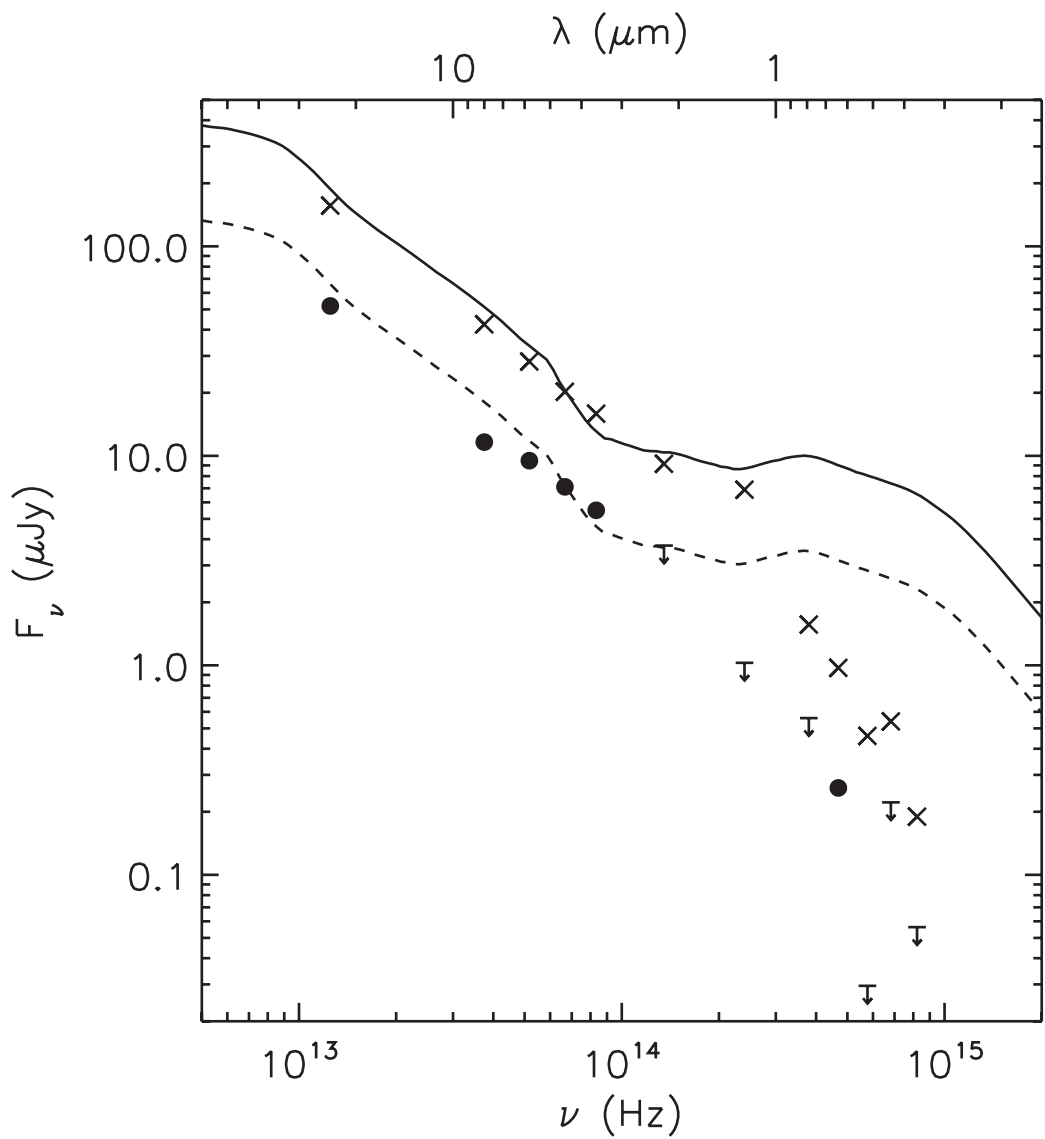}
\caption{Median observed SEDs for PL AGNs with and without X-ray
  detections.  The SEDs were constructed by taking the median observed flux
  density value at each band.   The crosses represent PL
  AGNs with X-ray counterparts, and
  the dots and upper limits represent PL AGNs without X-ray counterpart.    Upper
  limits are  $3\sigma$.  Templates are from \citet{elv94} redshifted
  to $z=1.71$, the median redshift of the X-ray sample, and
  normalized at 4.5~\micron.  The redshift chosen is somewhat
  arbitrary because only a small minority of sources have measured redshifts. }
\label{xraysedfig}
\end{figure}

PL AGNs with X-ray counterparts tend to have higher flux densities in
all wavebands than PL AGNs that are not X-ray sources.  In
particular, the X-ray sources are much more likely to have
corresponding visible light detections. (See Figure~\ref{xraysedfig}
and \S\ref{optical}.)  The greater rate of visible detections makes
X-ray-detected PL AGNs easier to measure spectroscopically than PL
AGNs lacking X-ray counterparts, and $1/3$ of the X-ray detected PL
AGNs (51 sources) have spectroscopic redshifts. The median redshift
is $z \approx 1.7$, and 80\% of these sources lie at $z > 1$.  In
contrast, only 3\% (11 sources) of the non-X-ray-detected PL AGNs
have measured redshifts.  These have a median $z \approx 1.2$, but
such a small and likely biased sample cannot yield firm
conclusions. The X-ray-detected PL AGNs were specifically targeted by
our MMT redshift survey, and therefore at least part of the higher
redshift detection rate for those sources is a selection effect, but
the difference in fluxes is real. The higher fluxes for X-ray sources
must be due to some combination of smaller distances, lower
obscurations, and/or higher intrinsic luminosities.  For example, the
process that fuels X-rays might also produce visible light.  The
overall similarity of the average SEDs (Fig.~\ref{xraysedfig}) tends
to suggest that there is little differential obscuration {\em affecting
the infrared emission\/}, though the fact that observed SEDs fall
below the AGN template in the visible and ultraviolet suggests that
obscuration is present in both classes.  However, infrared slopes do
not seem to  predict X-ray obscuration (\S\ref{hr}), and stacking of
undetected sources  (see \S\ref{stacking}) suggests that differential
obscuration is at least part of the story.  Nevertheless, greater distance
could also be important despite the few available redshifts  being low.  If
distance were the only factor, the X-ray-undetected sources would have to
have redshifts of order 2.5 (taking the K-correction into account) to
match the SED of the X-ray-detected sources.

\subsubsection{X-Ray Stacking}\label{stacking}

Stacking is a powerful way to examine the emission properties of
sources too faint to be identified individually \citep{geo08}.  There
are 330 PL AGNs that do not have strong X-ray counterparts in the
AEGIS-X catalog.  Removing 17 objects potentially contaminated by a
bright, nearby X-ray source leaves 313 stackable sources.  Stacking
X-ray counts (with the same methods used \citealt{geo08} including
the same 2\arcsec\ radius) yields significant detections in both the
soft and hard bands at a level $>$10$\sigma$.  The stacked soft band
flux is $(2.37 \pm 0.31) \times 10^{-17}$~erg~cm$^{-2}$~s$^{-1}$, and
the stacked hard band flux is $(1.35 \pm 0.20) \times
10^{-16}$~erg~cm$^{-2}$~s$^{-1}$. These values are about $1/3$ to
$1/2$ of the thresholds for individual detection.
In order to ensure that the stacked signal is not due to an unequal
contribution from a few X-ray sources that just missed being included
in the AEGIS-X catalog, we stacked again after removing the
weakly-detected X-ray sources.  There is still X-ray emission at
$5\sigma$ above background: soft band flux $(1.09 \pm 0.30) \times
10^{-17}$~erg~cm$^{-2}$~s$^{-1}$\ and hard band flux $(8.10 \pm 2.09)
\times 10^{-17}$~erg~cm$^{-2}$~s$^{-1}$. 
Thus the stacked emission appears to be
representative of the undetected population rather than dominated by
a few moderately bright X-ray sources, consistent with results in
fields with much deeper X-ray data \citep{alo06,don07}. 

The spectrum of the stacked signal is relatively hard with ${\rm HR} =0.05 \pm 0.10$.
After weakly-detected sources are removed, the spectral slope of the
remaining stack is likewise hard with ${\rm HR} = 0.18 \pm
0.18$. Assuming the average stacked source has $z=1.6$ (equal to the
median redshift of the PL AGN sample), the  hardness ratio
can be used to estimate obscuration by calculating column densities
using the PIMMS tool available from CIAO \citep[v3.9b,][]{muk93}.
The result is $N_{\rm H} \sim 10^{23}$~cm$^{-2}$ (assuming a
power-law model with intrinsic $\Gamma = 1.9$), implying a high but
not Compton-thick level of obscuration. The hardness ratio
is also consistent with a nearly flat X-ray spectrum, $\Gamma \sim1$, which
can be interpreted as evidence for a large fraction
of Compton-thick AGNs among the X-ray-undetected PL sources. In this
picture, the direct emission from the AGN is fully absorbed, and the
observed X-rays come from reflection off the torus or the accretion
disk \citep{gil07}. The reflected component has a very flat spectrum,
$\Gamma\la1$ \citep{George1991,Matt2000}, and does
not provide any information about the actual $N_{\rm H}$, just that
the source is likely to be Compton-thick. The HR alone cannot
discriminate between these possibilities or in general between
Compton-thick or Compton-thin obscuration. Moreover, sources below
but close to the Compton-thick limit may still be reflection-dominated in
the {\it Chandra} and XMM bands (depending on redshift), further complicating
the interpretation.  Whatever the explanation for the
hard spectrum may be, the hardness doesn't depend on whether or not the
weakly-detected sources are included in the stacking.

One possible way to discriminate between Compton-thick and
Compton-thin obscuration is to compare hard X-ray and infrared
fluxes. These two quantities are well correlated for AGNs with low to
moderate obscuration as first recognized by \citet{car87} and studied
in more detail by \citet{Lutz2004}. Figure~\ref{antonisfig} shows the
comparison for the PL AGN sample. The X-ray-detected objects fall
within or near the \citeauthor{Lutz2004} relation, even without
correcting the X-ray fluxes for any absorption. In contrast, the
stacking signal for the X-ray undetected PL AGNs is more than 100
times fainter than the relation, consistent with Compton-thick
absorption.  Stacking gives no information about individual objects,
but if some have relatively large X-ray fluxes and are therefore not
Compton-thick, the rest must have even smaller fluxes and be further
into the Compton-thick zone.  Thus it seems likely that many, perhaps
a majority, of the X-ray-undetected PL AGNs suffer Compton-thick obscuration.
\citet{Fiore2009}, using a different selection of infrared sources,
also found large numbers of X-ray-undetected, Compton-thick AGNs.

Another issue is the depth of the X-ray coverage.
\citet{geo08a} conducted an X-ray stacking analysis
of IRAC sources in the CDF-N, which
has even deeper IRAC data than the EGS.  They  found that
color-selected  sources  in the \citeauthor{ste05} AGN wedge produce
a soft spectrum consistent with $\Gamma \sim 
2.1$, whereas power-law sources produce a harder signal
($\Gamma \sim 1.6$) as expected of AGNs.  Our stacking analysis
(\S\ref{stacking}) shows that in the EGS, sources in
the \citeauthor{lac04}-defined wedge, sources in the
\citeauthor{ste05} wedge, and PL AGNs all  have similar stacked
fluxes and are all relatively hard with ${\rm HR} \approx-0.2$,
$-0.1$, and $0$ respectively.\footnote{Stacking for each wedge
  included all the IRAC  sources not detected as individual
  X-ray sources.  About  5\%  of the stacked sources in the
  \citeauthor{lac04}  and about $\sim$20\% in the 
  \citeauthor{ste05} wedge are  PL AGNs.  See \S\ref{color} for other
  statistics on the wedge populations.}  The
difference in results is likely to be due to X-ray depth of fields: more of
the faint X-ray sources are detected by the very deep exposures of
the CDF-N and excluded from the stacking analysis, whereas in the
shallower EGS, sources at similar X-ray fluxes remain
undetected and therefore are included in the stacked flux.  These
results suggest that many of the 
X-ray-undetected sources in the Lacy and Stern selection in the EGS
\emph{would} be detected with 2~Ms X-ray data.  However, the  soft spectra
of the fainter X-ray sources in the CDF-N suggests that the color-wedge-selected samples
at EGS-like depths
have a high rate of contamination from ULIRGs and other 
star-forming galaxies, as expected from the behavior of template
colors with redshift (\S\ref{color}).

\subsubsection{Comparison of X-ray and PL AGN Populations}

PL AGNs represent 22\% of the X-ray sources in the EGS.  This is much
higher than the 4\% PL AGNs in the general IRAC parent sample, but
sources with red IRAC power law SEDs are still a minority of X-ray
objects.  Relaxing the power-law definition to $P(\chi^2) \ge 0.01$
and $\alpha_{i} \le 0$ increases the fraction to 40\%, as found by
\citet{bar06} with that definition.  A comparable selection rate was
also found by \citet{car08} among their sample of X-ray sources with
$L_{\rm X} \ge 10^{42}$ erg~s$^{-1}$\ in the ECDFS using the definition
$\alpha_{i} \le 0$ and $P(\chi^2) \ge 0.05$.  Because most X-ray
sources in the EGS are expected to be AGNs,  requiring a red IRAC
power-law SED will not
identify even a majority of active nuclei.

The respective X-ray and PL AGN detection fractions can be estimated
under the assumption that the two methods are observing a single AGN
population and that detection by each method is {\em independent} of
the other.  Each method then detects a fixed fraction of the
population, and the product of the detection fractions times the
number in the population is the 159 X-ray-detected PL
AGNs.\footnote{\citet{Harwit} discussed the statistical method and
  assumptions.}  The resulting detection fractions are 0.33 for
X-ray, 0.22 for PL AGNs, and the population size is 2200 objects. The
assumption of a single population is probably wrong because the EGS
infrared observations reach higher redshifts than the X-ray
observations.  The effect of this error is that the X-ray population
is likely to be smaller and detection fraction larger than estimated
above.  Perfect independence of detection is also doubtful, despite
the most likely reasons for non-detection being entirely different
for the two methods (extinction for the X-rays, additional
contaminating radiation sources for the infrared), because higher
luminosity sources are more likely both to overcome extinction and to
outshine contaminating sources.  The effect of this error is that
both detection fractions are overestimated and the overall population
size is underestimated.  Thus the specific numbers above are at best
rough estimates, but the method is still useful for comparing AGN
populations found by different surveys.

Our study finds fewer X-ray sources with blue power law SEDs than
previous studies have, but this is at least partly because our power law
selection is more stringent.  According to our criteria, only 19\% of
X-ray sources fit a blue power law in the IRAC bands.  Relaxing the
criteria to those used by \citet{bar06} increases the fraction to
23\%, but this is still much smaller than the 38\% found by
\citeauthor{bar06} The discrepancy is mostly because the
observational error bars at 3.6 and 4.5~\micron\ are smaller in the
final catalog \citep{bar08} 
used here than in the earlier catalog used by
\citet{bar06}.\footnote{Systematic magnitude offsets between the two catalogs
  are only a few percent and thus insignificant.} \citet{car08} found
59\% of X-ray counterparts 
fit a blue power law in the IRAC bands, but they included sources
with detection significance as low as 3$\sigma$ instead of the
5$\sigma$ we required.  With large photometric error bars, a power
law will adequately fit many more sources.  Indeed
\citeauthor{car08} found 97\% of X-ray sources and 94\% of all IRAC
sources fit a power law versus our values of 41\% and 34\%
respectively.  Regardless of the exact numbers, the presence of
X-ray sources among this group means that a blue  SED, even one
closely fitting a power law in the IRAC bands, does
not rule out the presence of strong nuclear activity.

Requiring a power law fit from 3.6--24~\micron\ yields an even
smaller percentage of X-ray
objects: only 10\% of sources with 24~\micron\ and X-ray
detections fit a red power-law from 3.6 to 24~\micron\ (using our
original cut of $P(\chi^2) \ge 0.1$ and $\alpha_{m} \le -0.5$).  However,
while the X-ray sources with IRAC power laws are roughly evenly
divided between red and blue power-law sources, X-ray sources that fit
a 3.6--24~\micron\ power law are virtually all red.
(See Figure~\ref{alphacompfig}.)  Of the 70 3.6--24~\micron\
power-law sources with X-ray counterparts, only 5 exhibited the blue
power laws indicative of stellar emission.  This is hardly
surprising given the detection limits of the surveys; very few
sources will be detectable at 24~\micron\ unless the 24-\micron\ flux
density is considerably larger than the 8-\micron\ flux density.

\subsubsection{X-ray Luminosity}\label{xray_lum}

Rest-frame luminosity is a useful diagnostic of the power source
for an X-ray-emitting galaxy \citep{bar02}.
Rest-frame hard band (2--10~keV) X-ray luminosities were calculated
for the 47 sources with hard band detections and spectroscopic
redshifts using 
$L_{\rm X} = 4 \pi d_L^2 f_{\rm X} (1+z)^{\Gamma -2}$~erg~s$^{-1}$, where
$d_L$ is the luminosity distance calculated from Equation~1 of
\citet{pen99}, and $f_{\rm X}$ is the hard band flux in units of
erg~cm$^{-2}$~s$^{-1}$.  Most of these spectroscopically confirmed sources appear to
be obscured but not Compton thick (see \S\ref{hr} and Fig.~\ref{antonisfig}), so we
have assumed an intrinsic photon index of $\Gamma = 1.9$
\citep{nan94} for which the effects of absorption and K-correction
should be minimal \citep{bar02}.  All of these sources have X-ray
luminosities $L_{\rm X} > 10^{43}$~erg~s$^{-1}$\ with a maximum luminosity
of $L_{\rm X} \sim 10^{45}$~erg~s$^{-1}$.  Such high luminosities are
typical of AGNs.  This is further confirmation that our sample is not
likely to be dominated by star-formation, as $L_{\rm X}$ values of
local starburst systems typically lie below $10^{42}$ erg~s$^{-1}$\
\citep{bar02}. As expected \citep{alo06,don08}, 
virtually all of the highest X-ray luminosity sources are PL AGNs (see
Figure~\ref{lxhistfig}), and  35 out of the 50 sources with $L_{\rm
  X} > 10^{44}$~erg~s$^{-1}$\ are  PL AGNs.  Of the 15 exceptions,
12 have red-sloping IRAC SEDs but
were excluded by our goodness-of-fit criteria.  One source was
well fit to a blue power law with $\alpha_{i} \sim 0.3$, and two
sources were poorly fit to a blue power law with $-0.3 < \alpha_{i}
< -0.5$.  Thus, although only a minority (22\%) of X-ray sources are
PL AGNs, the majority (70\%) of the most X-ray-luminous objects are in
this class. The remaining 30\% of highly luminous X-ray sources
are likely AGNs  with 
contamination from stars and dust in the host galaxy preventing their
identification via the power-law selection.

A hard X-ray luminosity of $0.5-1\times10^{44}$~erg~s$^{-1}$\ marks a
division between PL AGNs and blue power-law sources
(Figure~\ref{lxhistfig}).  This luminosity is near the break ($L^*$) in
the X-ray luminosity function at our median sample redshift of $z
\sim 1.6$ \citep{ued03, sil08}.  PL AGNs having X-ray luminosities
below the break value are not excluded by low X-ray fluxes (as shown
by the stacking analysis) but rather by the absence of redshifts.
The high X-ray luminosity for galaxies detected as PL AGNs suggests
that the PL AGNs without X-ray counterparts are either very distant
compared to the sources with X-ray detections or else heavily
obscured.  Though the physical
interpretation behind the break in the luminosity function is still
unclear, the power-law selection may be finding an AGN population
that is distinct from the sub-$L^*$ X-ray galaxies.  The sub-$L^*$ X-ray
population may include AGNs with little dust near the nucleus or
different accretion modes (and hence lower intrinsic near-IR
emission) or AGNs  veiled by  emission from the host galaxy
(and hence not identified as an AGN by the near-IR SED).

Veiling by the host galaxy is probably at least part of the reason PL
AGNs with low luminosities are uncommon \citep{don07}.  Figure~\ref{lxlirfig} shows
the relation between hard X-ray flux and IR flux at different
wavelengths.  Among sources that exhibit both X-ray emission and PL
AGN SED shape, there is a strong linear correlation between X-ray and
IR flux.  This implies that PL AGNs are dominated by nuclear emission
(as demonstrated by \citealt{car87}), with both X-ray and IR
light closely related to the central engine.  In contrast, the
sources with IRAC and X-ray detections and spectroscopic
redshifts but that are not PL AGNs show much weaker correlation between IR and X-rays.
Indeed, the 3.6~\micron\ flux  for this sample is essentially constant,
suggesting that in the non-PL AGN sources, 3.6~\micron\ emission arises from a source
unrelated to the central engine.  Stellar emission from the host
galaxy is the obvious candidate.  In this picture, low-luminosity AGNs
are only identified as PL AGNs if they happen to reside in a
low-luminosity host galaxy.

\subsubsection{X-ray Hardness and Obscuration}\label{hr}

If the PL AGN sample preferentially selects sources that are obscured
at X-ray energies, one might expect the X-ray hardness ratio to be
greater for the PL AGN sample than the parent population.  However,
PL AGNs do not appear to exhibit significantly harder ratios than the
general sample (Figure~\ref{hrfig}).  One-third of the PL AGNs with
strong detections in both the hard and soft bands are considered hard
with ${\rm HR}> -0.2$ \citep[e.g.,][]{szo04}.  Of sources detected in
at least one band, the median ${\rm HR}=-0.30$, and 40\% have ${\rm
  HR}> -0.2$.  Weakly detected sources are harder with a median ${\rm
  HR}= 0.15$ and 72\% having ${\rm HR} > -0.2$.  The harder X-ray
spectra suggest that the X-ray-weak PL AGNs are more obscured than PL
AGNs with strong X-ray detections.  However, the correlation
coefficient between the IRAC power-law slope and the X-ray HR is
essentially zero, suggesting either little difference in obscuration
or that obscuration has little effect on the IRAC slope.  PL
selection seems to detect X-ray obscured and unobscured objects more
or less equally well, without a strong preference for one or the
other.  If this is so, the power-law selection technique can be a way
of identifying objects that are hidden behind gas and dust and
therefore missed by other surveys.

Figure~\ref{hrfig} shows little if any correlation between X-ray
hardness and infrared slope for PL AGNs.  Either the infrared slope
is controlled by properties other than extinction or the material
responsible for infrared extinction is not the same as the material
responsible for X-ray obscuration.

\begin{figure}
\includegraphics[trim=0 122 0 72, clip=t, width=6in]{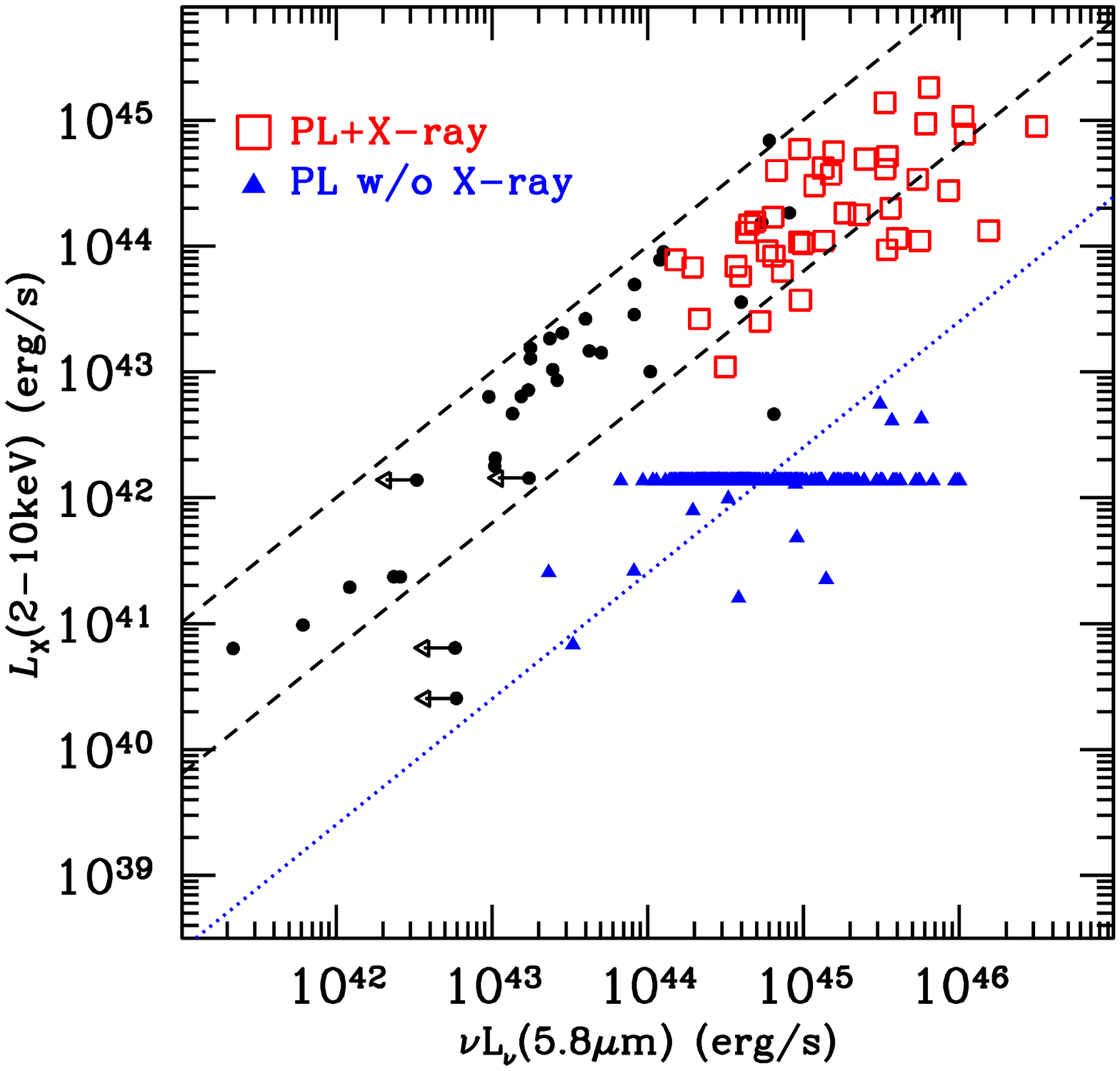}
\caption{
  X-ray 2--10~keV luminosity plotted against rest frame 5.8~\micron\
  luminosity. Filled circles (black) denote local AGNs from the sample of
  \citet{Lutz2004}.  The dashed lines show the dispersion around
  the mean $L_X({\rm 2-10~keV})$ to $\nu L_{\nu} (5.8\micron)$ relation
  for the \citeauthor{Lutz2004} sample, and the blue dotted line
  corresponds to X-ray luminosity 100 times lower than the average for
  their sample.   Our PL AGNs with X-ray detections and known redshifts are
  shown  with  red squares and those without X-ray detections with
  blue triangles.  These latter sources are assigned a flux density
  $8.1\times 10^{-17}$~erg~cm$^{-2}$~s$^{-1}$ (the stacking signal)
  and $z=1.6$ (median for the PL AGN sample) if no measured redshift.
  The factor of 100 is
  similar to the suppression of the observed X-ray emission of
  Compton-thick AGNs. In those sources 2--10~keV band is expected to be
  dominated by reflection, which is believed to represent
  1-2\% \citep{gil07} of the intrinsic AGN luminosity.}
\label{antonisfig}
\end{figure}

\begin{figure}
\includegraphics[trim=0 0 0 144,clip=t,width=6in]{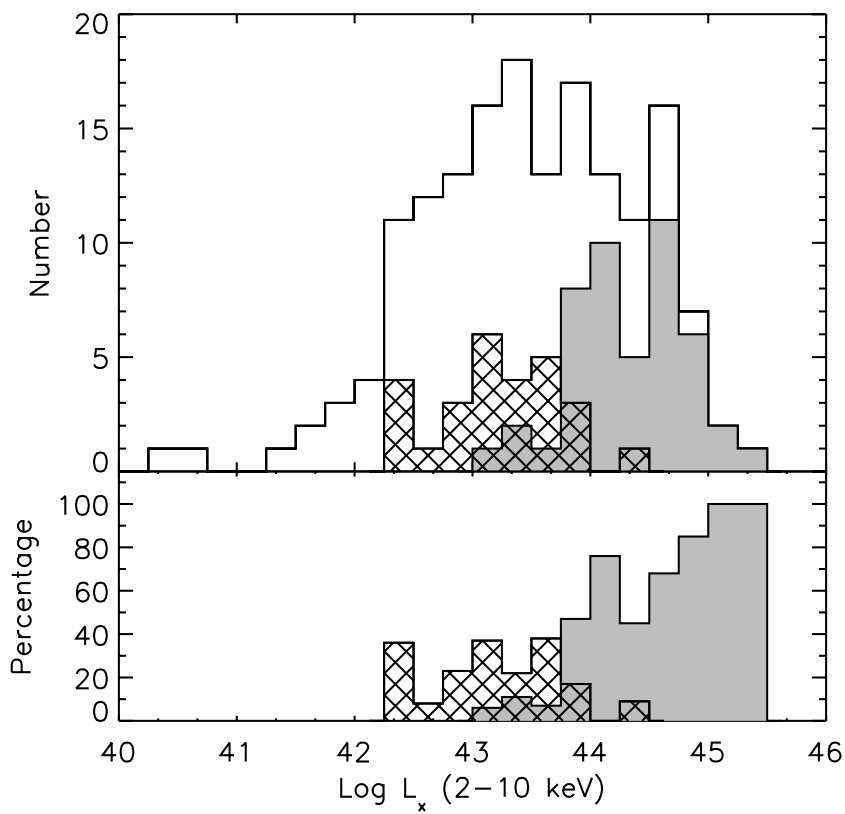}
\caption{ Histogram of hard band (2--10~keV) X-ray
  luminosities. In the 
  top panel, the empty histogram represents all X-ray sources with
  reliable redshifts,  the hatched histogram represents sources well fit
  to a blue power law in the IRAC bands, and the gray histogram
  represents the PL AGN sample, which make up the bulk of the
  high-$z$ X-ray population.  The bottom panel shows the fraction of
  X-ray sources detected as red (gray histogram) and blue (hatched histogram)
  power-law sources as a function of luminosity.}
\label{lxhistfig}
\end{figure}

\begin{figure}
\epsscale{0.8}
\plotone{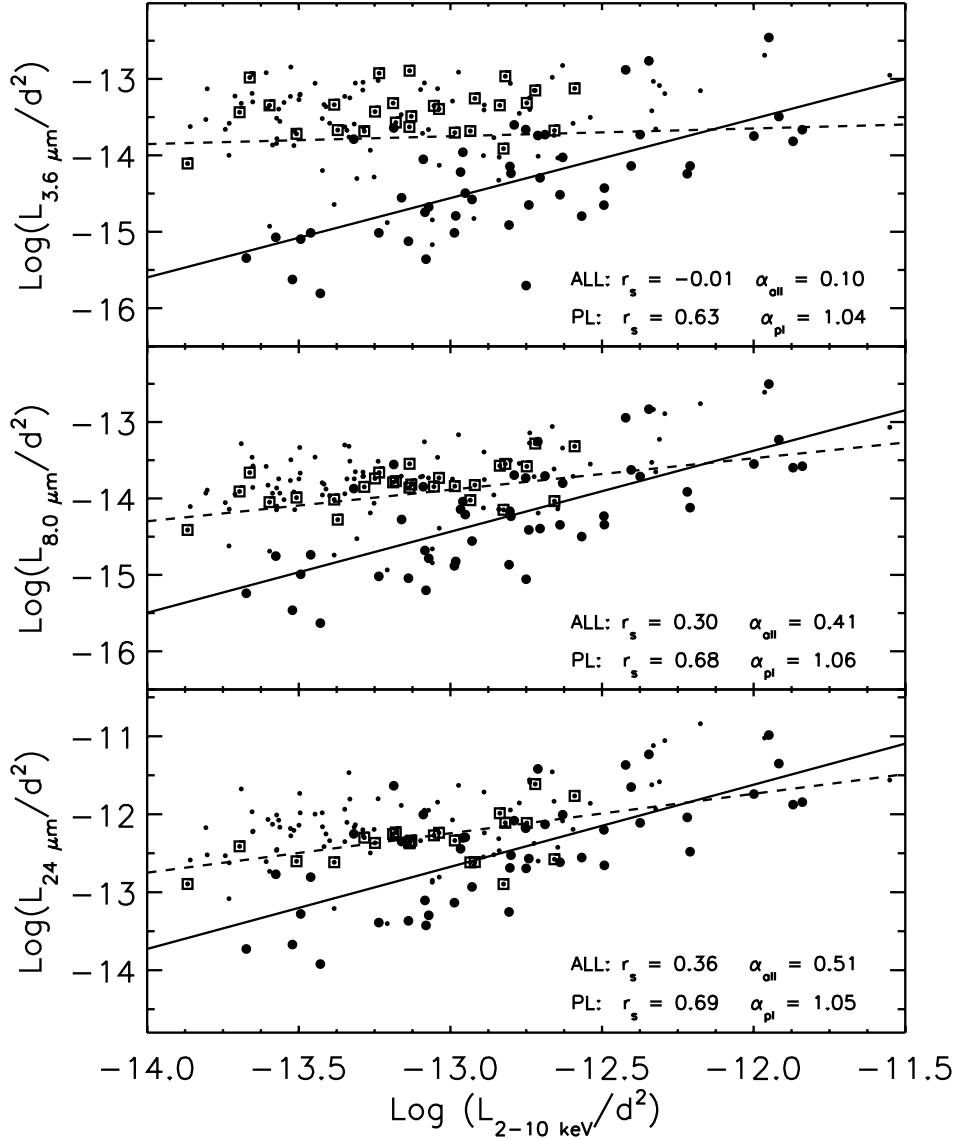}
\caption{Rest frame infrared fluxes versus rest frame X-ray flux.
  Small dots represent
  all X-ray sources with infrared counterparts and known
  redshifts. Open squares
  surround X-ray sources well fit to a blue power law, and
  large filled circles denote the X-ray-detected PL AGNs.  The solid lines
  indicate the best fits through the PL AGN sample, and the
  dashed lines denote the best fits through the parent sample.
  Luminosities include the respective K-corrections and
  have been divided by the luminosity  
  distance squared in order to avoid non-physical correlations caused
  by correlation of all luminosities with distance.  Units are erg~s$^{-1}$~cm$^{-2}$.
  Correlation coefficients and the slopes of the best fit lines are
  given for the parent sample (ALL) and the red PL AGN sample (PL).}
\label{lxlirfig}
\end{figure}

\begin{figure}
\includegraphics[trim=0 0 72 216,clip=t,width=6in]{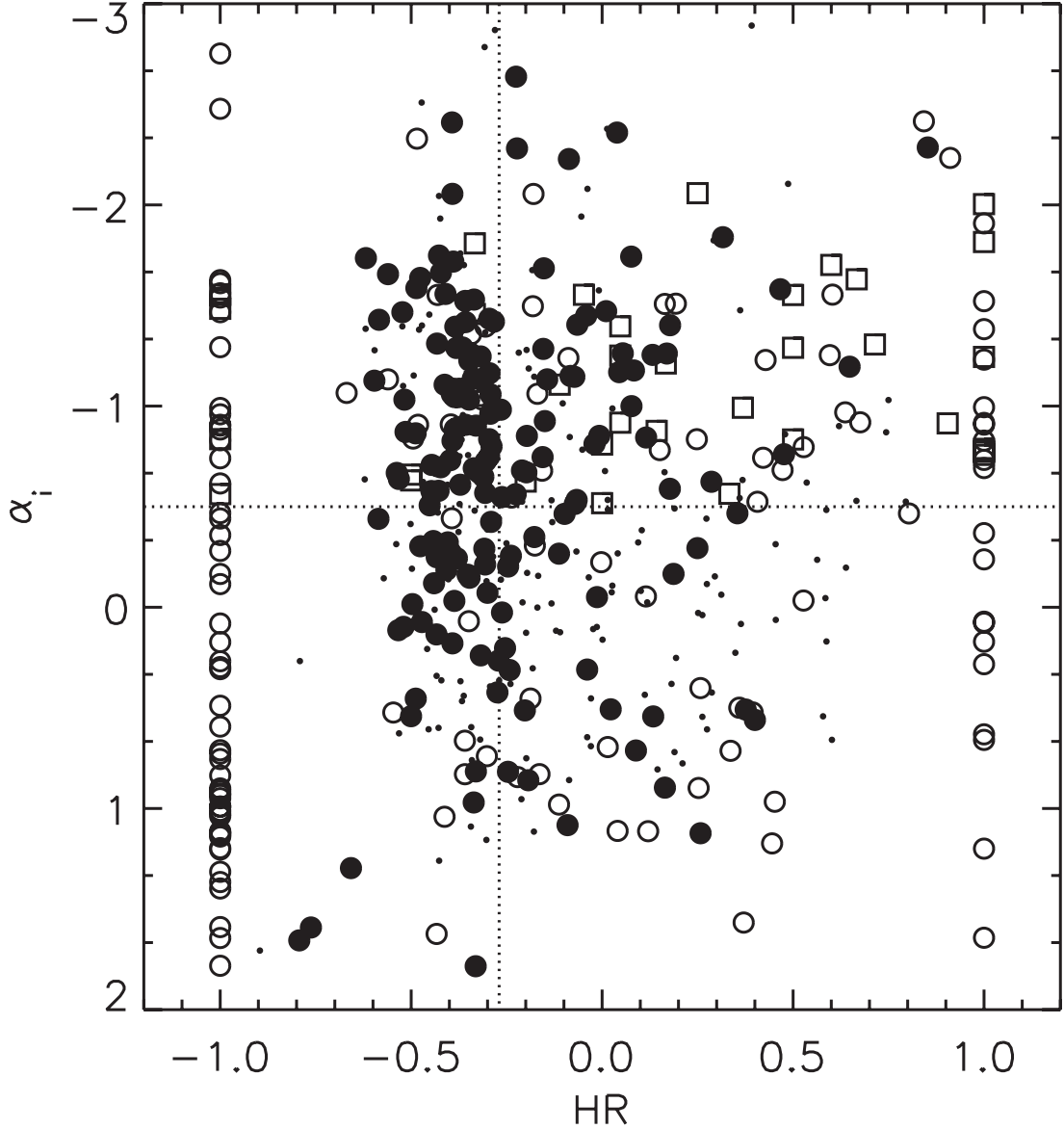}
\caption{Hardness ratios for all IRAC/X-ray sources versus the 3.6 to
  8.0~\micron\ IRAC slope.  Small dots represent IRAC sources with
  X-ray counterparts that could not be fit to a power-law; large
  symbols represent sources that are well fit to a power law.  Filled
  circles indicate sources with strong hard and soft band detections
  (Poisson probability $\!< 4 \times 10^{-6}$), while open circles
  represent sources strongly detected in either the hard or soft band
  and weakly detected in the other band.  (Open circles plotted at
  ${\rm HR}<0$ represent sources strongly detected in the soft band,
  and open circles at ${\rm HR}>0$ represent sources strongly
  detected in the hard band.) Open squares represent PL AGNs
  that are weakly detected (but with Poisson spurious
  probability $<\!2 \times 10^{-2}$) in both X-ray bands.  The horizontal line indicates
  the power-law cut-off for our selection; sources with $\alpha_{i} <
  -0.5$ (above the line) are PL AGNs.  The vertical line indicates
  the median HR value for all of the strongly detected IRAC/X-ray
  sources.  }
\label{hrfig}
\end{figure}

There are 46 PL AGNs with spectroscopic redshifts that have been
strongly detected in both the hard and soft bands for which the 
column density can be estimated using PIMMS.  Of these, 90\% are
obscured with column densities $N_{\rm H} > 10^{22}$~cm$^{-2}$.  Two sources
(EGSPL~80/153) have hardness ratios consistent with being
Compton thick ($N_{\rm H} > 10^{24}$~cm$^{-2}$), but the ratios of
X-ray to 5.8~\micron\ flux (among the red squares in
Fig.~\ref{antonisfig}) indicate less obscuration than this.  Probably these
sources have high but not
Compton-thick obscuration, perhaps with a harder-than-average intrinsic
spectrum.  Another atypical source
could not be fit using $\Gamma = 1.9$ and appears to require a
steeper intrinsic slope to get the observed counts.  All column
density  values here are
uncertain because errors in hardness ratios can be large, and reliable
estimates of obscuration levels will require proper
spectral fitting analysis \citep{lai06}.  Indeed
the simple obscuring slab geometry 
assumed in the calculations may not
apply at all \citep{LaMassa2009}.  Nevertheless, the hardness ratios and
column density calculations suggest a high amount of obscuration among
the X-ray sources, but none of the {\em detected} sources is
likely to be Compton thick (Fig.~\ref{antonisfig}).

\subsection{Infrared Properties}\label{ir}

\subsubsection{24--70~\protect\micron\ Flux Densities}

Among the PL AGN sample, 399 (82\%) have 24~\micron\ counterparts
within 1\farcs5.  Some of the PL AGNs may have escaped 24~\micron\
detection because of the 9.7~\micron\ silicate absorption feature,
which is commonly seen in absorbed AGNs \citep{rie75,roc91}.  The
absorption band will be redshifted to 24~\micron\ at $z \sim 1.5$,
near our median redshift value.

\begin{figure}
\plotone{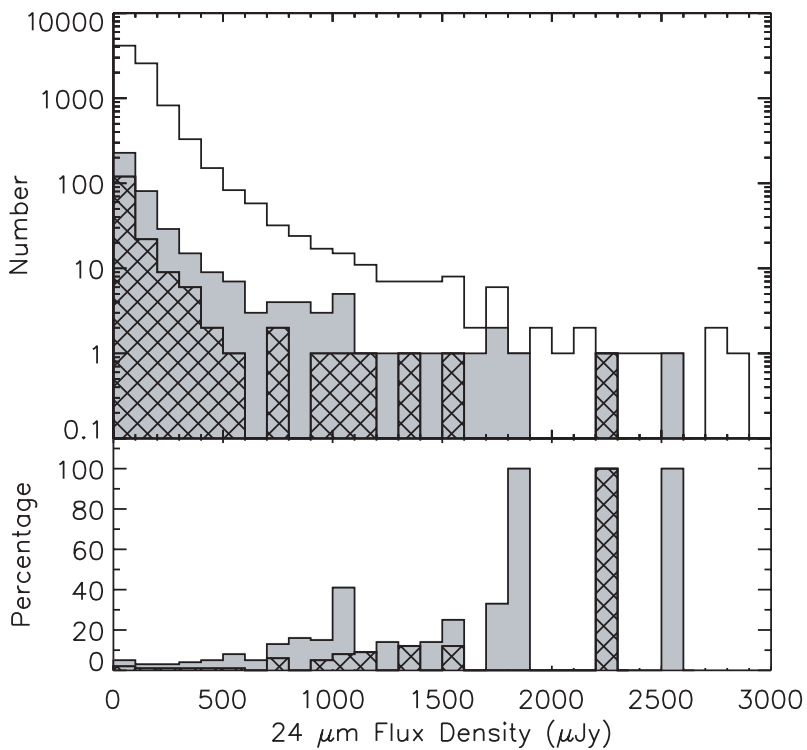}
\caption{Distribution of 24~\micron\ flux densities.  In
  the top panel, the empty histogram represents all IRAC sources with
  24~\micron\ detections plotted on a logarithmic scale. The gray
  histogram represents sources selected as PL AGNs, and the hatched
  histogram represents objects with 3.6 to 24~\micron\ red power laws
  ($\alpha_m<-0.5$).
  The bottom panel shows the percentage of PL AGNs (gray) or
  3.6--24~\micron\ red power-law sources (hatched)
  in the IRAC/24~\micron\  parent sample in each flux density bin.  
  }
\label{mipsfluxfig}
\end{figure}

Many IR luminous galaxies are thought to host AGNs, and \citet{wee06}
and \citet{bra06} demonstrated that the AGN fraction in
24~\micron-selected samples tends to increase as 24~\micron\ flux
density increases from 0.1 to 1~mJy.  Yet 57\% of the PL AGNs with
24~\micron\ counterparts have $S_{24} < 100$~$\mu$Jy, and only
4\% have flux densities exceeding 1~mJy (see Fig.~\ref{mipsfluxfig}
for the distribution of flux densities).  The median 24~\micron\ flux
density of the PL AGN sample is in fact smaller than the median of
the entire IRAC/MIPS population because the power-law selection
excludes strong star forming systems, which are bright in the mid-IR.
In agreement with \citet{don07}, the 24~\micron\ detection fraction
is high (100\%) for objects with X-ray luminosity $L_X \approx
10^{41}$~erg~s$^{-1}$.  The X-ray emission in these relatively weak
X-ray sources
is likely to come from stellar processes rather than an AGN.  Near $L_X
\approx 10^{42}$~erg~s$^{-1}$, the 24-\micron\ detection fraction drops to
$\sim$80\%, but it  rises again to 100\% with increasing X-ray
luminosity.  Thus 24~\micron\ is not a particularly efficient
wavelength to search for AGN emission, but AGNs generally will emit
at that wavelength in accordance with the power-law
character of their SEDs \citep{elv94}.  The relative depths of the
X-ray and 24-\micron\ surveys in the EGS mean nearly all X-ray AGNs
are detected at 24~\micron.

Extending the IRAC power law to 24~\micron\ is not a good predictor
of actual 24~\micron\ flux density for individual objects, as noted
by \citet{bar06}.  Only 11\% of PL AGNs with 24~\micron\ detections
have actual 24~\micron\ flux density within 10\% of the value
predicted by extrapolating the IRAC power law.  In our sample, the
extrapolated 24~\micron\ flux density is greater than the actual flux
density 40\% of the time and less than the actual flux density 60\%
of the time.  While the power law character of the SED is a good
approximation statistically,  actual SEDs show breaks in the
(observed) 3.6--24~\micron\ range as would be found for local QSOs
\citep{Netzer2007} if they were observed at redshifts characteristic
of the PL AGN sample.

Only eight PL AGNs have a detection at 70~\micron. These
sources have atypically high $S_{24}/S_{70}$ ratios compared
to the parent sample, but this SED is consistent with the expected AGN
energy distributions \citep{elv94}.

\subsubsection{Variability}

Because AGNs are so compact, they can vary on short timescales.
\citet{Koz10} found typical AGN
variability  at IRAC wavelengths to be about 0.05~mag in six months,
though of course individual objects can vary more or less than this amount.
The IRAC EGS data were taken in two epochs: 2003 December and 2004
June--July \citep{bar08}, and 
we have analyzed the two epochs  separately
to check for flux variability.
Variability was defined such
that $|\Delta S| / \delta S > 4$, where $\Delta S$ is the difference
in flux density between the two epochs and $\delta S$ is the
quadrature sum of the individual measurement uncertainties.  The
probability of random observational errors causing a constant source
to vary by 4$\sigma$ in one IRAC bandpass is $6 \times 10^{-5}$.  In
the full sample of 11787 IRAC sources, less than one source would be
expected to show spurious variation at this level in one bandpass and
a negligible fraction in all four bandpasses.

Among our 489 PL AGNs,
397 were separately detected in both the epoch 1 and epoch 2
catalogs.  Two sources ($\sim$1\%) were reliably observed to vary in the
mid-IR by $>$4$\sigma$ in all bandpasses: EGSPL~122 increased in flux
density at all four wavelengths from 2003 to 2004, while EGSPL~153
decreased at all wavelengths.  Both sources have flux densities
$\ga$100~$\mu$Jy in the four IRAC bandpasses (AB magnitudes 17--18),
15--60 times higher than the
median flux densities for the PL AGN sample.
The actual variation over the 6 months amounted to a difference of
about 10\%, easily detectable with the large $S/N$.  
Both sources  have X-ray counterparts, as expected for bright AGN.

A third source (EGSPL~35) showed more complex brightness variations. 
This source has IRAC flux densities 400--700~$\mu$Jy
(AB magnitude 16--17). It also varied by $>$4$\sigma$ across all four IRAC bandpasses, but it
increased its flux density at 3.6~\micron\ and 5.8~\micron\ while
decreasing at 4.5~\micron\ and 8.0~\micron.  (There are no obvious
image artifacts near this or any of the variable sources in the PL AGN
sample.)  IRAC observes wavelengths 3.6~\micron\ and 5.8~\micron\
simultaneously and wavelengths 4.5~\micron\ and 8.0~\micron\
simultaneously.  In 2003 December, approximately 21 hours passed
between the 3.6/5.8~\micron\ observations and the 4.5/8.0~\micron\
observations of EGSPL~35.  Thus, it is possible that these
observations may have caught a brief infrared flare.

Lowering our variability threshold to $3\sigma$ results in only one additional
PL AGN varying across all four channels.  However, $\sim$51\% of our PL AGNs
varied by $3\sigma$ in at least one IRAC channel, and $\sim 32$\%
varied by $4\sigma$ in at least one channel.  The high incidence of
apparent variability is further confirmation of the AGN character of
the PL AGN sample.

Though the number of variable PL AGNs is too few to make statistical
comparisons to the parent sample, there were 15 sources (including
the two PL AGNs) in the IRAC population with consistently increasing
or decreasing flux densities across all four IRAC channels between
the two epochs.  Forty percent of these are identified as AGNs via
their X-ray emission and are thus likely to be real variables. The
rate of detecting a non-X-ray  source that varies in the IR is therefore
$\la$0.1\% in our data, suggesting that photometric failures causing
spurious variability are minimal.  Thus variability appears to be
real and characteristic of AGNs, as found by \citet{Koz10}.  As also
shown by those authors, even four epochs are not sufficient to find a
majority of the variable population, and two epochs as here will be
even less complete.


\subsection{Visible Light Properties}\label{optical}

\begin{figure}
\plotone{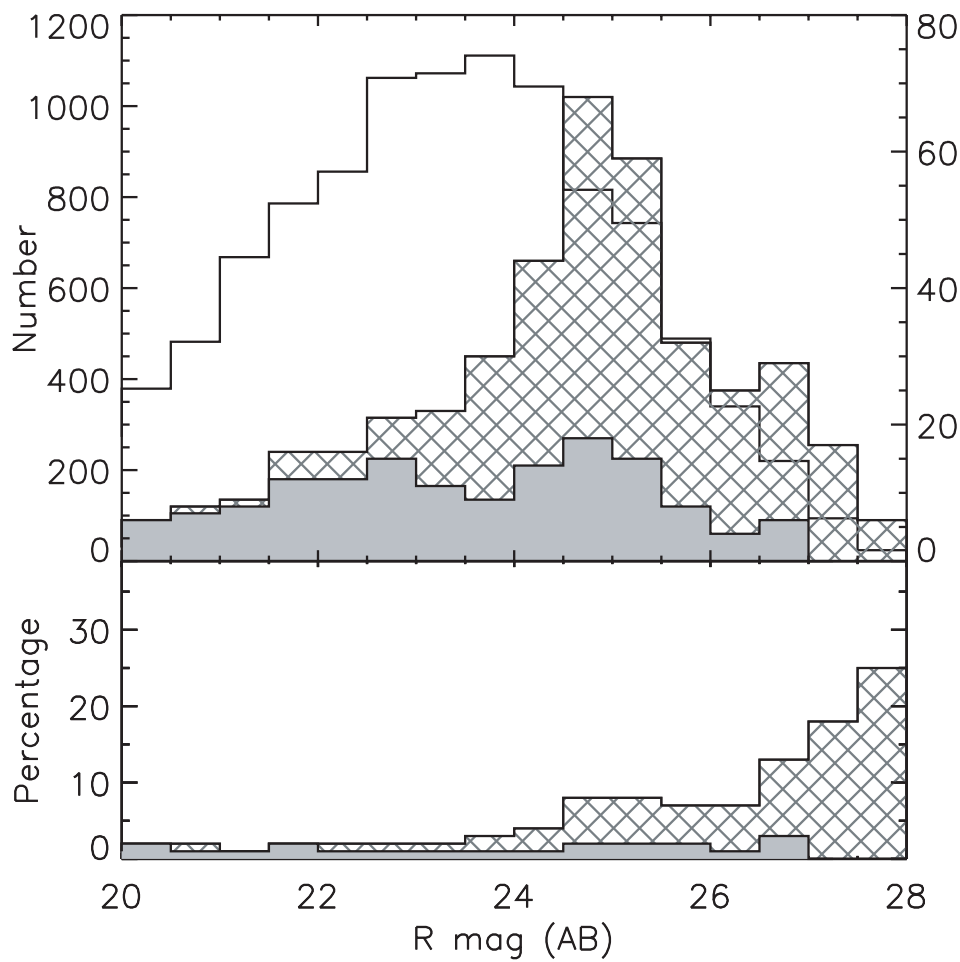}
\caption{Histogram of $R$-band magnitudes for sources detected at $R$.  In the top
  panel, the white histogram represents all  IRAC sources (with
  5$\sigma$ detections in all IRAC bands); values are indicated by the scale on the
  left.  The hatched histogram denotes PL AGNs, and the gray histogram
  denotes PL AGNs with X-ray detections, both  with values indicated
   by the scale on the right.  The bottom panel shows
  the percentage of PL AGNs  (hatched) and the percentage of PL AGNs with X-ray
  detections (gray) at each
  $R$-band magnitude in the IRAC parent sample.  All of the
  histograms exclude sources 
  undetected at $R$, which are 9\% of the IRAC sources and 18\% of
  the PL AGNs.}
\label{rbandhistfig}
\end{figure}

PL AGNs have a lower rate of visible light detection compared to the
parent sample of
IRAC sources, and the detections tend to
be fainter (median $R_{AB}$ = 24.5 versus 23.0).\footnote{As
  throughout this paper, the IRAC parent sample includes only sources
  detected in all four bands with $S/N>5$.}
The relative faintness in visible light is expected if
the IRAC power-law extends even roughly  into the
near-IR and visible bands, as many  do (see 
\S\ref{sed}).  Eighteen percent of our sample is optically faint and
not detected by Subaru down to a limit of $R \sim 26.5$ at $5 \sigma$
or CFHT down to a limit of $R \approx 24.2$ at $8 \sigma$ as compared
to 9\% of the parent IRAC sample being undetected at $R$.
Figure~\ref{rbandhistfig} shows the distribution of $R$-band magnitudes
for the PL AGN sample.  

\begin{figure}
\epsscale{0.8}
\plotone{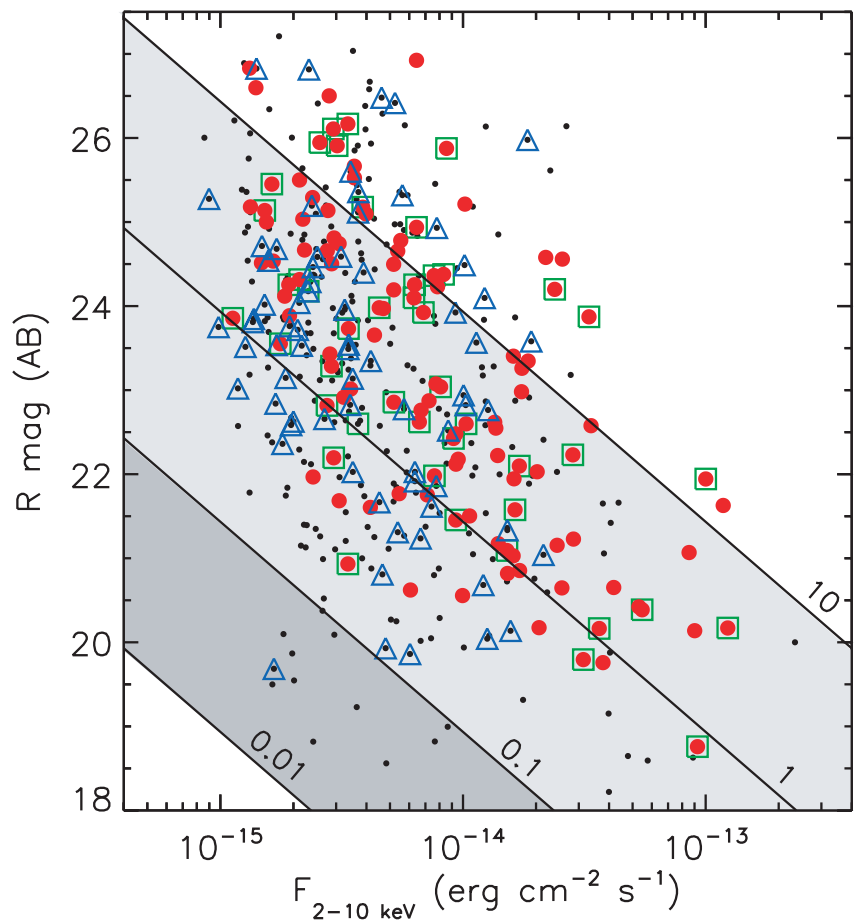}
\caption{Hard X-ray flux versus  $R$ band
  magnitude.  Small black dots indicate all IRAC sources with X-ray and $R$
  band counterparts.  Large red dots represent PL AGNs,
  green squares represent sources that have power laws from 3.6 to
  24~\micron, and blue triangles indicate sources that are well 
  fit to a blue IRAC power law.  The numbers near the slanted lines indicate
  the $X/O$ ratio of the line.  The light shaded area denotes the
  region where most AGNs are expected to lie; the
  darker shaded area indicates the transition region that is populated
  by both AGNs and starbursts.}
\label{rxfig}
\end{figure}

\begin{figure}
\plotone{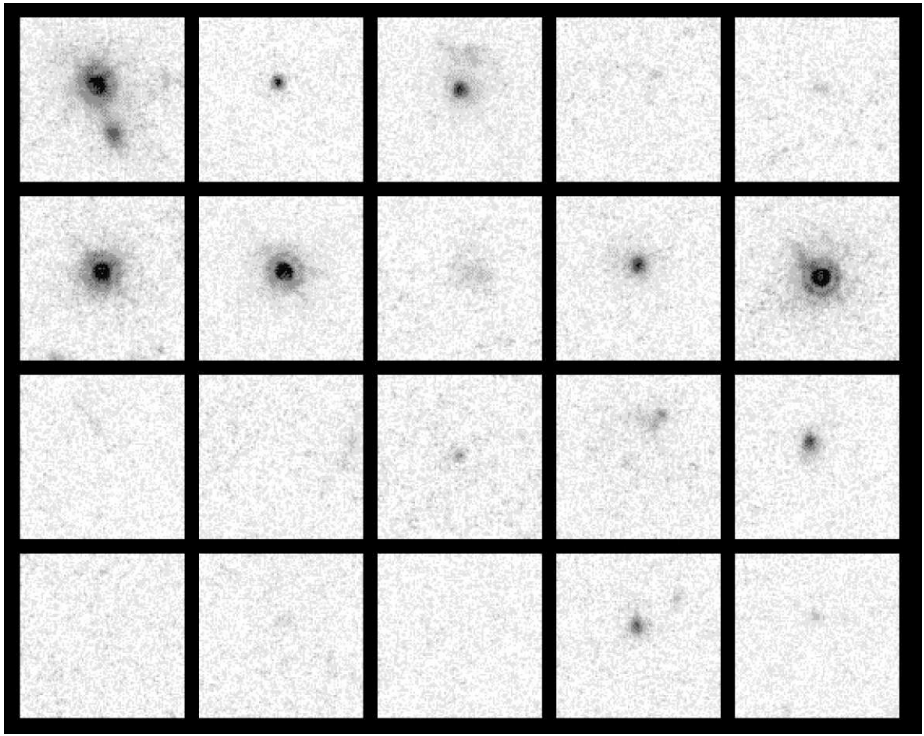}
\caption{ACS I negative images for a sample of PL AGNs.  The top
  two rows depict sources with X-ray detections, and the bottom two
  rows show sources without X-ray counterparts.  Thumbnails are
  8\arcsec\ on each side.}
\label{morphfig}
\end{figure}

X-ray to visible (``optical'') ratios ($X/O$) are commonly used to separate AGNs from
starburst galaxies \citep{mac88, ale01, hor01, bar03} and are defined
as $X/O = (R / 2.5) + 5.5 + \log(F_{\rm HB})$ where $R$ is the
$R$-band magnitude (Vega) and $F_{\rm HB}$ is the X-ray hard band
flux.  Because AGNs are powerful enough to fuel energetic X-rays, they
are expected to have high $X/O$, 0.1--10 or greater.  $X/O \sim
0.01$--0.1 generally indicates low luminosity AGNs or galaxies
dominated by star formation, and even lower $X/O$ is characteristic of normal
galaxies, which have very weak hard X-ray emission.
At the depth of our data, nearly all detected X-ray sources are expected
to be AGNs, and indeed 95\% of the X-ray plus IRAC sample have $X/O >
0.1$.
There are 116 PL AGNs that have both optical $R$ band detections and
strong X-ray hard band detections; this is 95\% of all hard band
detected PL AGNs.  All of these have $X/O$  consistent with values
expected for AGNs (see Figure~\ref{rxfig}).  The X-ray stacked source
also lies in the AGN region, though outside the limits of the
plot.  
In general, PL AGNs with X-ray detections are optically brighter than
sources without X-ray detections (see Figure~\ref{morphfig}).
This is consistent with the X-ray-detected population being at lower
redshift, less obscured, or both.

Nearly a quarter of the PL AGNs with both $R$ and X-ray detections have extreme
X-ray to optical ratios of $X/O > 10$ (Fig~\ref{rxfig}).
In many of these, the ratio is large because the objects
are quite faint in $R$ band rather than because they have
stronger X-ray fluxes than average.  Objects with large $X/O$ ratios are
candidates for being at high redshift and  Compton thick
\citep{civ05}, though Compton-thick seems unlikely for any of the
X-ray-detected objects because of the
observed X-ray to infrared ratios (Fig.~\ref{antonisfig}).
Sources with $X/O > 10$ have a median hardness ratio of ${\rm HR}\sim 0$,
compared to the median of ${\rm HR}= -0.3$ among all of the X-ray-detected
PL AGNs.  Less than half of these sources have available {\it HST}/ACS
observations, and of these most are too faint to determine morphology
by eye.  However, three sources  appear extended while one
appears point-like.  Beyond their identification as AGNs, the nature
of the extreme $X/O$ sources remains in doubt.

\citet{pie07} observed the host galaxy morphologies of IRAC power-law
objects with $\alpha_{i} < 0$ within the redshift range $0.1 < z <
1.2$ and found that X-ray selected AGNs lie mostly in E/S0/Sa
hosts and not in  Sc/Sd/Irr galaxies.  Conversely, PL AGNs without
X-ray detection inhabit host galaxies of all types, though the
selection  may
work against AGNs hosted by E/S0/Sa galaxies.  Because X-ray undetected
PL AGNs are optically very faint, it is difficult to obtain
morphologies for these sources, and those lacking redshifts were
not included in the \citet{pie07} sample at all.
However, Figure~\ref{morphfig} suggests that X-ray-detected PL
AGNs and X-ray undetected PL AGNs may inhabit morphologically
different galaxies.  Undetected PL AGNs  may be  surrounded by
dust and gas, which obscure both the 
visible light and the X-ray emission.  Emission from the dust could
contribute to emission at 24~\micron\ and longer wavelengths.
The morphological difference suggests that obscuration is at least
part of the reason for many PL AGNs not being detected in X-rays, but
higher redshifts could well be important too.

\subsection{Radio Properties}

\begin{figure}
\epsscale{0.8}

\plotone{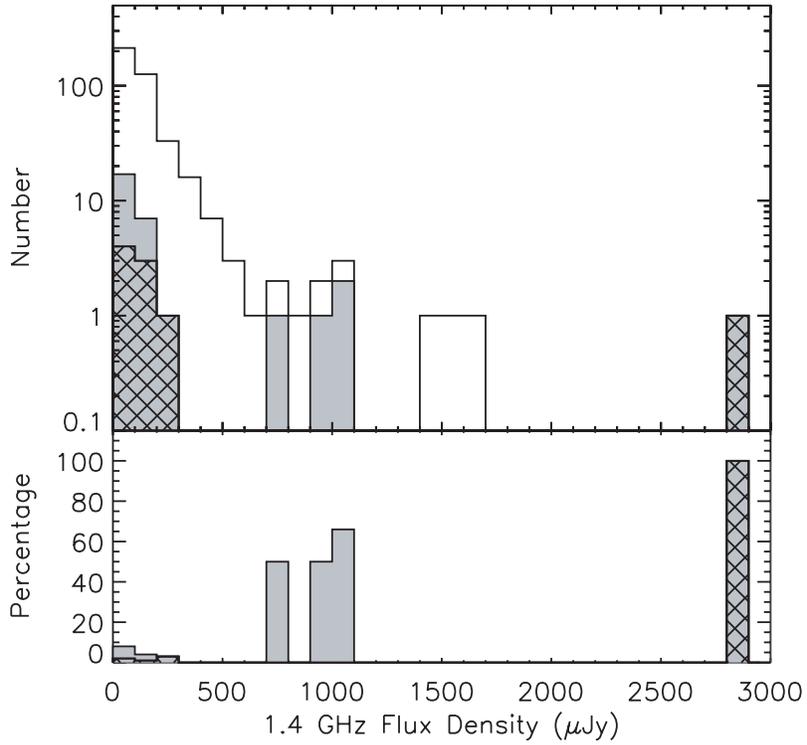}
\caption{Distribution of 1.4~GHz flux densities.  In the
  top panel, the empty histogram denotes all IRAC sources with 1.4~GHz
  detections.  The gray histogram represents  PL AGNs,
  and the hatched histogram shows sources that have red power
  laws from 3.6 to 24~\micron.  The bottom panel shows the percentage
  of radio sources at each radio flux density that are PL AGNs (gray) or have 3.6--24~\micron\ red
  power law SEDs (hatched).}
\label{radiofluxfig}
\end{figure}

The VLA observations cover roughly two-thirds of the IRAC $+$ {\it Chandra\/}\
field.  Of the 357 PL AGNs  within the radio field and
outside of high-noise regions, 31 ($\sim$9\%) have 1.4~GHz
counterparts with  $S/N> 5$.  Over half of these
have flux densities $S_{1.4 \rm~GHz} < 100$~$\mu$Jy, and over 75\%
have $S_{1.4 \rm~GHz} < 200$~$\mu$Jy (see Figure~\ref{radiofluxfig}).
Four sources (13\%) have $S_{1.4 \rm~GHz} > 1$~mJy.  This is 
higher than the $\sim$5\% of all radio sources that have 
$S_{1.4 \rm~GHz} > 1$~mJy, though barely statistically significant.
However, 1/4 of galaxies with $S_{1.4 \rm~GHz} > 1$~mJy are PL AGNs.
One source (EGSPL~260) is very radio-bright with $S_{1.4
\rm~GHz} \approx 10$~mJy. This source is considerably
brighter in all observed wavebands than the respective median flux densities:
it has  $S_{24} \approx 1.25$~mJy and 
an X-ray detection that is $\sim$10 times higher than
the median X-ray flux observed for {\it Chandra\/}\ sources (but near the
median X-ray flux for the PL AGN sample).  ACS morphology shows that
the source is point-like, but no redshift has yet been measured for this
object.

The ratio of radio to 24~\micron\ flux densities can be used  as a
sign of AGN emission.  All but one of the PL AGNs with radio
counterparts also have detections in the 24~\micron\ band, allowing
us to study the ratio between the radio and mid-IR flux densities
using $q\equiv \log(S_{24} /S_{1.4\rm~GHz})$. Star-forming systems show a strong
correlation between radio and far infrared flux densities
\citep{con92}, and this correlation extends to 24~\micron\
\citep{app04}. \citet{Rieke2009} discussed measurements of $q$ for
local star-forming galaxies and found $q\ga1.22$, but the value of
$q$ at a fixed observed wavelength depends on redshift.  Lacking
redshifts for most of the PL AGNs, we stick to observed flux densities rather
than attempt uncertain K corrections, and the most suitable
comparison is $q = 0.84 \pm 0.28$ \citep{app04}.  Sources with $q <
0$ have strong radio emission relative to their 24~\micron\ flux
densities and are considered radio-loud AGNs \citep{don05,
  par08}. Eleven out of the 30 PL AGNs (37\%) with both radio and
24~\micron\ detections have $q < 0$.  This is somewhat higher than
the 20\% of sources in the EGS radio/IRAC sample that have $q < 0$,
and also higher than the fraction found by \citet{don07}, who found
that only 2 out of their 18 power-law sources had $q < 0$.  However,
the small-number statistics and different radio depths of the EGS and
GOODS fields make any detailed comparison problematic.  An additional
ten PL AGNs have $0<q<0.56$, outside the range expected for star
formation but not quite in the radio-loud category.  The remaining 9
radio/24\micron/PL AGNs have $q$ values within the expected range for
star-forming galaxies and non-radio-excess AGN with the largest value
being $q=1.26$.  

Only four of the radio-detected PL AGNs have redshifts, and
luminosities have been calculated for these sources assuming a radio
spectral index of $\alpha = 0.8$ \citep{yun01}.  Two sources had
$L_{1.4 \rm~GHz} > 10^{24} $~W~Hz$^{-1}$ at redshifts of 1.4 and 1.7,
while two had $L_{1.4 \rm~GHz} \sim 2$--$4 \times 10^{23}$~W~Hz$^{-1}$
at redshifts of 0.5 and 1.  These luminosities are consistent with
values observed for AGN radio sources in the local universe, which
typically have $L_{\rm radio} > 10^{23}$ W Hz$^{-1}$
\citep{yun01}. IRAC sources that are not PL AGNs but do have radio
counterparts have a median redshift of $z= 0.6$ and median luminosity
of $L_{1.4 \rm~GHz} \sim 10^{23} $~W~Hz$^{-1}$.  These could be
either AGNs or luminous starbursts.

A stacking analysis of the PL AGNs without radio detections yields
significant detection ($S/N>20$) with a median 1.4 GHz flux density
of $13.5 \pm 0.7~\mu$Jy per beam after accounting for bandwidth
smearing, stacking losses, and stacking bias.
Assuming the stacked source is at the median
redshift $z=1.6$, it would have a luminosity of $L_{1.4 \rm~GHz} =
1.9 \times 10^{23}$ W Hz$^{-1}$, well within the AGN range
\citep[e.g.,][]{yun01}. If this stacked source were powered by star
formation, its luminosity would translate to a star formation rate of
$>$200~${\rm M}_{\odot}$~yr$^{-1}$ for a Salpeter IMF.  Thus explaining
the radio emission for much of the sample by star formation requires
a high star formation rate; an AGN attribution seems more likely.
A $q$ value for the
stacked source can be estimated using the median 24~\micron\ flux
density for the PL AGN sample, 86~$\mu$Jy, giving $q=0.8$ and
implying that few if any of the radio-undetected sources can be
radio-loud AGNs.  Thus the radio-loud fraction of the
PL AGN population is 11 $q<0$ detected sources out of 357 PL AGNs in the
radio area, ${\lesssim}5$\%.  This fraction  is typical of
other AGN surveys \citep[e.g.,][]{Green2009}\footnote{Determining the fraction of the
  AGN population 
  that is radio loud is complex and controversial, and the result
  likely depends on luminosity and redshift.  \citet{Jiang2007}
  and \citet{Rafter2009} provide recent overviews.}
and  again consistent with the AGN character of the PL AGN sample.

\begin{figure}
\plotone{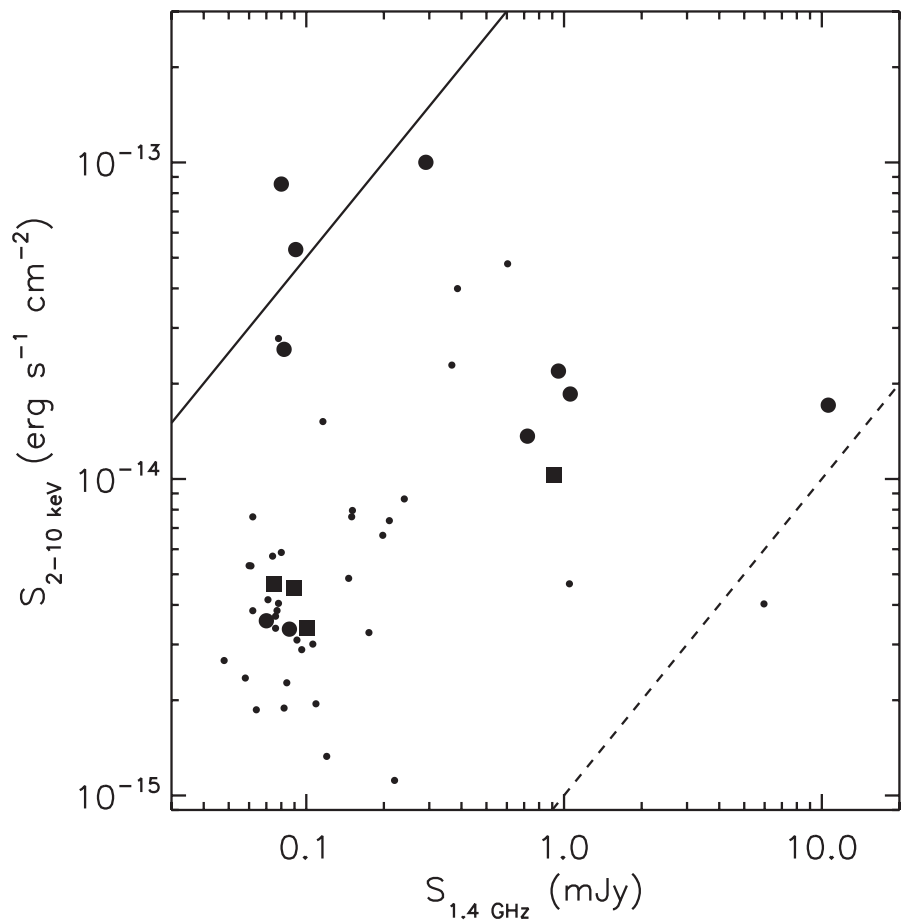}
\caption{Radio flux density versus hard band X-ray flux.
  Small dots denote all IRAC sources with X-ray and radio detections.
  Filled circles represent  PL AGNs, and filled
  squares represent objects with a blue power law from
  3.6 to 8.0~\micron.  The solid line indicates the X-ray/radio
  correlation \citep{Brinkmann2000} for 
  radio-quiet AGNs and is given by $S_{2-10 \rm~keV}$ (W m$^{-2}$) $=
  10^{-15.3}S_{1.4 \rm~GHz}$ (mJy).  The dashed line indicates the
  correlation for starburst galaxies, with $S_{2-10 \rm~keV}$ (W
  m$^{-2}$) $= 10^{-18.0}S_{1.4 \rm~GHz}$ (mJy). Numerical values for
  both correlations
  come from  \citet[their equations 7 and 9]{sim06}. }
\label{radxfig}
\end{figure}

Ten of the 30 PL AGN radio sources have X-ray counterparts. This
fraction is consistent with expectations from a study of $q$-selected
sources in the general IRAC/X--ray/radio source population
\citep{par08}.  EGSPL~260, the bright radio source mentioned above,
has an X-ray counterpart, but among the rest of the sources in the
small sample, the presence of X-ray counterparts does not appear to
be correlated with the radio flux densities, as shown in
Figure~\ref{radxfig}.  The PL AGN sample in general lies closer to
the AGN correlation than do other X-ray and radio sources, although
there is one PL AGN source near the starburst correlation line. (This
is again EGSPL~260.) The majority of the sources lie between $100 <
S_{1.4 \rm~GHz} < 300$~$\mu$Jy, and these modest radio sources tend
to show X-ray emission above the starburst correlation. The
clustering of sources at low radio flux densities and low X-ray
fluxes may be an indication of X-ray absorption and the limits of our
survey rather than an inherent attribute of these sources.  As
\citet{sim06} noted, the starburst and AGN correlations shown in 
Figure~\ref{radxfig} may be overestimates; they are derived by using a relatively
deep radio survey and a relatively shallow X-ray survey.  Hence, the
true correlation may actually lie lower than plotted and include more
of the population than currently indicated. This would be consistent
with the X-ray emission from PL AGNs being primarily produced by the
nucleus, rather than star formation, as also argued in
\S\ref{xray_lum}.

\section{SUMMARY}

The IRAC power-law selection technique is designed to identify
sources that are dominated by AGN emission in the infrared bands
regardless of obscuration.  Only 4\% of the IRAC sources and 2\% of
the IRAC/MIPS sources in the EGS are identified as power-law AGNs
using our criteria.  PL AGNs have SEDs that generally follow AGN
templates with either a rough power-law shape from IRAC to visible
wavelengths or a downturn in flux density at the visible wavelengths.
This may be due to gas and dust obscuring the visible light from PL
AGNs.  PL AGNs by definition exhibit no SED `bumps'.  This is unlike
much of the rest of the IRAC parent population, which often show
clear stellar emission features in their SEDs.

PL AGNs typically have low visible flux densities but are highly
luminous overall.   Most of the luminous X-ray sources
($L_{\rm x} \ge 10^{43}$ erg~s$^{-1}$) in the EGS are PL AGNs.
The high luminosity of PL AGNs allows their
detection  at high redshifts, and indeed the EGS PL AGNs lie
at a median redshift of $z \sim 1.6$, much higher than the median
redshift of $z \sim 0.7$ of the parent IRAC population.

One-third of the PL AGN sample have X-ray counterparts, a detection
rate greater than the 6\% found for the parent population.  The
detection fraction rises slightly if low-significance X-ray
detections are included.  However, among all X-ray-detected IRAC
sources, only about one-fifth exhibit red power-law SEDs, while a
similar fraction exhibit blue power-law SEDs (with spectral index
$\alpha_{i} > -0.5$); the rest are not well fit by a power law.  As
most of the X-ray sources in our field are expected to be AGNs, the
lack of a red power law in the IRAC bands is not an indication of
inactivity in a galaxy.  Rather, the power-law method finds
energetically dominant AGNs and leaves out sources with host galaxy
stellar contributions rivaling or exceeding the AGN infrared
light. There is a clear division between red power-law sources and
blue power-law sources near the break luminosity in the X-ray
luminosity function at $L_{X} \sim 10^{44}$ erg~s$^{-1}$, suggesting
a physical transition from sources dominated by host galaxy light to
those dominated by nuclear emission.  While fewer than half the PL
AGNs are invididually detectable in X-rays with 200~ks {\it
  Chandra\/}\ exposures, an X-ray stacking analysis found a
significant X-ray signal for the rest.  Among the X-ray-detected
sources with secure redshifts, most of the sources are obscured with
$N_H> 10^{22}$~cm$^{-2}$, but none is likely to be Compton-thick with
$N_H > 10^{24}$~cm$^{-2}$.  Assuming the stacked source has a
redshift equal to the median, the observed hardness ratio corresponds
to a column density of $10^{23}$~cm$^{-2}$, but the ratio of
mid-infrared to X-ray fluxes for the stacked source implies that a
substantial fraction of the undetected sources are Compton thick.

Eighty-two percent of the PL AGN sample have 24~\micron\
counterparts.  Though the AGN fraction in other samples increases with
higher 24~\micron\ flux densities, the majority of our PL AGNs are
not especially mid-IR luminous with 57\% of them having $S_{24} <100~\mu$Jy.  
The  IRAC power law cannot be extended to
predict observed 24~\micron\ flux density --- only 11\% of PL AGNs have
IRAC power laws that extrapolate to 24~\micron\ flux densities  within
10\% of the actual values.

In two epochs separated by six months, there are only mild
indications of infrared flux variability in the PL AGN sample. While about one-third
of the PL AGNs varied by at least $4\sigma$ in at least one IRAC band, only a few objects
exhibited variability exceeding $4\sigma$ across all four IRAC
bandpasses. These results are consistent with other studies of AGN variability,
but the number of epochs is too small to make definitive conclusions
about infrared variability in particular.

Though the power-law selection tends to identify optically-faint
galaxies, 82\% of the PL AGNs have $R$-band detections.  All sources
with both $R$-band and hard X-ray detections  have
X-ray/optical ratios of 0.1 or greater, consistent with values
expected for AGNs.  Nine percent of the PL AGNs have 1.4~GHz
counterparts.  Half of these have $S_{1.4 \rm~GHz} < 100~\mu$Jy, 
suggesting that PL AGNs are not preferentially radio-bright.
Nearly forty percent of the PL AGNs with radio counterparts have 
$q\equiv\log(S_{24} /S_{1.4\rm~GHz}) < 0$. These sources
have radio-to-mid-infrared flux ratios higher than typical for
star-forming galaxies and can be characterized as `radio-loud';
however they make up only a small fraction of the overall PL AGN sample.

Ultimately, to uncover the mysteries of galaxy formation and
evolution, a complete and reliable census of AGNs must be obtained
for study.  To develop such a census, we must discover what types of
sources are identified by all of the various AGN selection
techniques.  The power-law selection method returns highly luminous,
often obscured, nuclear-dominated galaxies at intermediate to high
redshifts.  The method is not intended to select all AGNs or even all
infrared AGNs; a rough estimate of the selection fraction is 22\%.
Sources with dominant contributions from stellar emission, though
they may comprise a large fraction of the total AGN population, will
be excluded from this survey.  However, the power-law technique is
intended to reliably select active sources with little contamination,
and in particular can select heavily obscured AGNs (and hence AGNs
missed by X-ray, visible, or radio surveys). Thus, it provides an
excellent way of supplementing other AGN selection methods and leads
to a more complete AGN census for study.

\acknowledgements 

We thank Mark Dickinson for supplying FIDEL data in advance of
publication and Alison Coil and Christopher Willmer for advice
concerning the spectroscopic data.
This study has made use of data from AEGIS, a
multiwavelength sky survey conducted with the Chandra, GALEX, Hubble,
Spitzer, Keck, Palomar, CFHT, MMT, Subaru, VLA, and other telescopes
and supported in part by the NSF and NASA.  This work is based in part
on observations made with the Spitzer Space Telescope, which is
operated by the Jet Propulsion Laboratory, California Institute of
Technology under a contract with NASA. Support for this work was
provided by NASA.  This work is based in part on data collected at
Subaru Telescope, which is operated by the National Astronomical
Observatory of Japan and on observations obtained with
MegaPrime/MegaCam, a joint project of CFHT and CEA/DAPNIA, at the
Canada-France-Hawaii Telescope (CFHT) which is operated by the
National Research Council (NRC) of Canada, the Institut National des
Sciences de l'Univers of the Centre National de la Recherche
Scientifique (CNRS) of France, and the University of Hawaii. This work
is based in part on data products produced at TERAPIX and the Canadian
Astronomy Data Centre as part of the Canada-France-Hawaii Telescope
Legacy Survey, a collaborative project of NRC and CNRS.
The observations reported here were obtained in part at the MMT 
Observatory, a facility operated jointly by the Smithsonian Institution 
and the University of Arizona. PB acknowledges research support
from the Natural Sciences and Engineering Research Council of Canada.

Facilities: \facility{CFHT}, \facility{CXO}, \facility{Keck:II},
\facility{Spitzer (IRAC, MIPS)}, \facility{Subaru (Suprime)},
\facility{VLA}, \facility{MMT (Megacam, Hectospec)}


\bibliographystyle{apj}

\clearpage


\clearpage

\clearpage

\begin{deluxetable}{ccccrrrr}
\tabletypesize{\scriptsize}
\tablecaption{IRAC Sources and Categories of Power-Law Fits\label{tbl-1}}
\tablewidth{0pt}
\tablehead{
\colhead{Sample (IRAC Sources)} &  \colhead{Fit Range} &
\colhead{Red PL\tablenotemark{a}}      &  \colhead{Red $\alpha$\tablenotemark{b}}   &
\colhead{Red non-PL\tablenotemark{c}}  &   
\colhead{Blue PL\tablenotemark{a}}     &  \colhead{Blue $\alpha$\tablenotemark{b}} &
\colhead{Blue non-PL\tablenotemark{c}}
}
\startdata
All  &
$3.6-8.0$ \micron\  & 489 (4\%) & $-1.03$ &156 (10\%) & 3352 (30\%)  & 0.97 & 6182 (55\%)  \\
With 24~\micron\ counterparts &
$3.6-8.0$ \micron\  & 399 (2\%) & $-1.05$ & 1078 (14\%) & 1774 (22\%) & 1.62 & 4727 (59\%) \\
With 24~\micron\ counterparts &
$3.6-24$ \micron\  & 168 (5\%) & $-1.11$ & 6523 (82\%) & 41 (0.5\%) & 1.68 & 1247 (16\%)  \\
Red 3.6--24~\micron\ power law\tablenotemark{d} &
$3.6-8.0$ \micron\  & 153 (91\%) & $-1.09$ & 12 (7\%) & 3 (2\%) & $-0.35$ &0 (0\%)  \\
With X-ray counterparts &
$3.6-8.0$ \micron\  & 159 (22\%) & $-1.09$ & 136 (19\%) & 137 (19\%) & 0.50 & 287 (40\%)  \\
\enddata
\tablenotetext{a}{Number and percentage of sources fit to a power law
  with $P_x \ge 0.1$. ``All'' refers to sources detected at
  $\ge$5$\sigma$ in all four IRAC bands.
  `Red' refers to sources with spectral index $\alpha \le -0.5$; `Blue'
  refers to sources with $\alpha > -0.5$.}
\tablenotetext{b}{Median spectral slope}
\tablenotetext{c}{Percentage of red-sloping or blue-sloping sources
  poorly fit to a   power-law}
\tablenotetext{d}{Sources that  fit a red power law
  in the 3.6--24~\micron\ range}
 
\end{deluxetable}

\clearpage

\begin{deluxetable}{cccccccccccccc}
\rotate
\tabletypesize{\scriptsize}
\tablecaption{Properties of Power-Law AGNs\label{tbl-2}}    
\tablewidth{0pt}
\tablehead{
\colhead{EGSPL}         &  \colhead{EGSIRAC\tablenotemark{a}}         &  
\colhead{$\alpha_i$\tablenotemark{b}}     &  \colhead{$P_i$\tablenotemark{c}}          &  
\colhead{$\alpha_m$}       &  \colhead{$P_m$}              &  \colhead{$z$}           &  
\colhead{$S_{3.6}$} &  \colhead{$S_{4.5}$} &  
\colhead{$S_{5.8}$} &  \colhead{$S_{8.0}$} &  
\colhead{$S_{24}$}  &  \colhead{[LNG2009]-egs\tablenotemark{d}}
&  \colhead{$\log L_X$\tablenotemark{e}}
\\
\colhead{}              &  \colhead{}              &  
\colhead{3.6--8 } &  \colhead{3.6--8 \micron} &  
\colhead{3.6--24 }&  \colhead{3.6--24 \micron}&  \colhead{}      &
\colhead{$\mu$Jy}       &  \colhead{$\mu$Jy}       &  
\colhead{$\mu$Jy}       &  \colhead{$\mu$Jy}       &  
\colhead{$\mu$Jy}       &  \colhead{}              &  \colhead{erg~s$^{-1}$}
}
\startdata
1  &    J141427.18+520139.6  &  -1.12  &  0.29  &  -1.39  &  0.03  &  \nodata  &  10.15  &  13.78  &  19.71  &  24.04  &  142.10  &    \nodata  &  \nodata  \\
2  &    J141428.02+520352.8  &  -1.15  &  0.26  &  -1.15  &  0.44  &  \nodata  &  7.11  &  10.35  &  12.84  &  17.78  &  66.16  &    \nodata  &  \nodata  \\
3  &    J141428.13+520347.1  &  -1.54  &  0.62  &  -1.23  &  0.00  &  \nodata  &  17.59  &  23.33  &  36.54  &  59.05  &  185.77  &   0008  &  \nodata  \\
4  &    J141431.06+520358.6  &  -0.62  &  0.21  &  -1.16  &  0.00  &  \nodata  &  29.59  &  35.54  &  36.38  &  50.01  &  236.47  &   0010  &  \nodata  \\
5  &    J141433.18+520255.0  &  -2.06  &  0.73  &  -1.48  &  0.00  &  0.77  &  131.39  &  199.25  &  338.60  &  675.07  &  2550.26  &    \nodata  &  \nodata  \\
6  &    J141438.02+520315.4  &  -1.53  &  0.86  &  -1.53  &  0.96  &  \nodata  &  4.72  &  6.73  &  10.56  &  15.79  &  86.10  &    \nodata  &  \nodata  \\
7  &    J141442.37+520241.5  &  -0.56  &  0.40  &  \nodata  &  \nodata  &  \nodata  &  7.09  &  8.42  &  10.68  &  10.06  &  \nodata  &   0023  &  \nodata  \\
8  &    J141443.94+520040.1  &  -1.39  &  0.65  &  -1.37  &  0.83  &  \nodata  &  6.64  &  9.55  &  12.20  &  20.51  &  90.65  &    \nodata  &  \nodata  \\
9  &    J141446.03+520627.7  &  -1.21  &  0.20  &  -1.65  &  0.00  &  \nodata  &  4.71  &  5.69  &  9.81  &  11.72  &  99.86  &    \nodata  &  \nodata  \\
10  &    J141446.13+520614.4  &  -0.73  &  0.11  &  -0.82  &  0.14  &  \nodata  &  14.24  &  17.82  &  23.34  &  24.81  &  70.11  &    \nodata  &  \nodata  \\
\enddata

\tablecomments{
IRAC flux densities in this table were measured in 3\arcsec\ diameter
apertures and are the ones used in this paper.  They
are 3\% higher than the final published values \citep{bar08}, which
should be used for future work.
Table~\ref{tbl-2} is published in its entirety in the electronic
edition of the Astrophysical Journal.  A portion is shown here for
guidance regarding its form and content.
}
\tablenotetext{a}{IRAC IDs  from the catalog published by \citet{bar08}.}
\tablenotetext{b}{Calculated IRAC power-law spectral index ($S_{\nu}
  \propto \nu^{\alpha}$);  $\alpha \le -0.5$ is required for
  selection as a PL  AGN.} 
\tablenotetext{c}{Goodness of fit expressed as a
  probability.  The numerical value is the probability that
  observations of a source
  with a power-law SED would give the actual chi-square or larger. 
  $P_x > 0.1$ is required for selection as a PL  AGN.} 
\tablenotetext{d}{X--ray IDs from the catalog published by \citet{lai08}.}
\tablenotetext{e}{X--ray luminosity in the 2--10~keV band}
\end{deluxetable}

\begin{table}
\scriptsize
\begin{center}
\caption{\protect\centering X-ray Detection Rates\tablenotemark{a}\label{tab-3}}
\begin{tabular}{lcll}
\hline\hline
Sample & Detection Rate & Comment&Reference\\
\hline
EGS PL AGNs  &33\%       &200~ks, $P({\rm false})<4 \times 10^{-6}$&this paper, \citet{lai08}\\
EGS PL AGNs  &40\%       &200~ks, $P({\rm false})<2 \times 10^{-2}$&this paper\\
EGS 3.6--24~\micron\ PL sources &39\%&200~ks, $P({\rm false})<4 \times 10^{-6}$&this paper, \citet{lai08}\\
All IRAC sources &\06\%&200~ks, $P({\rm false})<4 \times 10^{-6}$&this paper, \citet{lai08}\\
IRAC sources with 24~\micron\ counterpart &\07\%&200~ks, $P({\rm false})<4 \times 10^{-6}$&this paper, \citet{lai08}\\
CDF South PL AGNs &53\%     & 1~Ms data&\citet{alo06}\\
CDF North PL AGNs &55\%   &$P({\rm false})< 1\times 10^{-7}$&\citet{don07}\\
CDF North PL AGNs &85\%   &$P({\rm false})< 6 \times 10^{-3}$&\citet{don07}\\
\hline
\end{tabular}
\end{center}
\tablenotetext{a}{Fraction of sample detected in X-rays with
  specified observation and detection criteria}
\end{table}

\end{document}